\AtBeginDocument{}
\documentclass[aps,prl,twocolumn,superscriptaddress,calc,epsfig,10pt,floatfix]{revtex4-1}
\setcitestyle{super} % makes citations superscript with natbib

\usepackage{graphicx}
\usepackage{graphics}
\usepackage{amssymb,amsmath}
\usepackage{wasysym}
\usepackage{float}
\usepackage{physics}
\usepackage{color}
\usepackage[normalem]{ulem} 
\usepackage[pdftex,breaklinks,colorlinks=true,linkcolor=blue,citecolor=blue,urlcolor=blue]{hyperref}
\usepackage{tikz}
\usetikzlibrary{calc,shapes}
\usepackage{lineno}  % Makes linenumbers work well. Have to manually adjust \internallinenumbers right above caption call for figures

\newcommand{\mitCUAaddress}{Department of Physics, MIT-Harvard Center for Ultracold Atoms, and Research Laboratory of Electronics, MIT, Cambridge, Massachusetts 02139, USA}

\begin{document}

\title{Quantum Register of Fermion Pairs}

\author{Thomas Hartke, Botond Oreg, Ningyuan Jia, and Martin Zwierlein}
\date{\today}
\affiliation{\mitCUAaddress}

\begin{abstract}
   Fermions are the building blocks of matter, forming atoms and nuclei, complex materials and neutron stars. Our understanding of many-fermion systems is however limited, as classical computers are often insufficient to handle the intricate interplay of the Pauli principle with strong interactions. Quantum simulators based on ultracold fermionic atoms instead directly realize paradigmatic Fermi systems~\cite{Inguscio2008,Bloch2008Many,Zwerger2012,Inguscio2016,Gross2017QSim}, albeit in ``analog'' fashion, without coherent control of individual fermions. Digital qubit-based quantum computation of fermion models, on the other hand, faces significant challenges in implementing fermionic anti-symmetrization~\cite{Abrams1997Simulation, Bravyi2002Fermionic,Bauer2020Quantum}, calling for an architecture that natively employs fermions as the fundamental unit. Here we demonstrate a robust quantum register composed of hundreds of fermionic atom pairs trapped in an optical lattice. With each fermion pair forming a spin-singlet, the qubit is realized as a set of near-degenerate, symmetry-protected two-particle wavefunctions describing common and relative motion. Degeneracy is lifted by the atomic recoil energy, only dependent on mass and lattice wavelength, thereby rendering two-fermion motional qubits insensitive against noise of the confining potential. We observe quantum coherence beyond ten seconds. Universal control is provided by modulating interactions between the atoms. Via state-dependent, coherent conversion of free atom pairs into tightly bound molecules, we tune the speed of motional entanglement over three orders of magnitude, yielding $10^4$ Ramsey oscillations within the coherence time. For site-resolved motional state readout, fermion pairs are coherently split into a double well, creating entangled Bell pairs. The methods presented here open the door towards fully programmable quantum simulation and digital quantum computation based on fermions.
\end{abstract}

\maketitle

The Pauli principle lends stability to matter, from the shell structure of nuclei and the periodic system of elements to Pauli pressure protecting a neutron star from gravitational collapse. In search of a robust quantum information architecture, one may strive to emulate the particular stability of noble gases and magic nuclei, provided by fully-filled shells of fermions. Interactions between fermions can further enhance stability, e.g.~via formation of Cooper pairs in nuclear matter and superconductors, opening energy gaps and thereby carving out protected subspaces for fermion pairs. Thus fermion anti-symmetry and strong interactions, the core challenges for classical computations of many-fermion behavior~\cite{Bauer2020Quantum}, may offer decisive solutions for protecting and processing quantum information~\cite{Kitaev2001Unpaired}.

Recent advances in quantum gas microscopy of bosonic~\cite{Bakr2009, Sherson2010} and fermionic~\cite{Cheuk2015Quantum,Haller2015Single,Parsons2015,Omran2015,Edge2015,Brown2017Spin,Koepsell2020,Hartke2020,Gross2020} atoms, including optical tweezer approaches~\cite{Murmann2015, Labuhn2016Tunable,Bernien2017Probing}, have enabled analog quantum simulations of fermionic systems at the resolution of individual fermionic atoms~\cite{Mazurenko2017Antiferro,Brown2019Badmetal,Nichols2019spintransport,Koepsell2019Polaron,Gall2021competing}. These platforms thus offer clear prospects to experimentally realize digital quantum computing with fermionic hardware.

Here we demonstrate an array of fermion pairs for robust encoding and manipulation of quantum information. We leverage fermionic statistics to initialize~\cite{Viverit2004, Serwane2011, Chiu2018} a low-entropy array of spin-singlet fermion pairs, and to restrict the fermion dynamics to a symmetry-protected subspace.
Quantum information is encoded in the motional state of the atom pair by forming superpositions of center-of-mass and relative vibrational motion.
Working in a subspace of pairs of atoms decouples the quantum information from environmental noise~\cite{Kwiat2000, Kielpinski2001}.
Strong interactions are induced via a Feshbach resonance, allowing for tunability of the two-fermion motional qubit frequency over several orders of magnitude.

\begin{figure}[!t]
	\centering
	\includegraphics[width=\columnwidth]{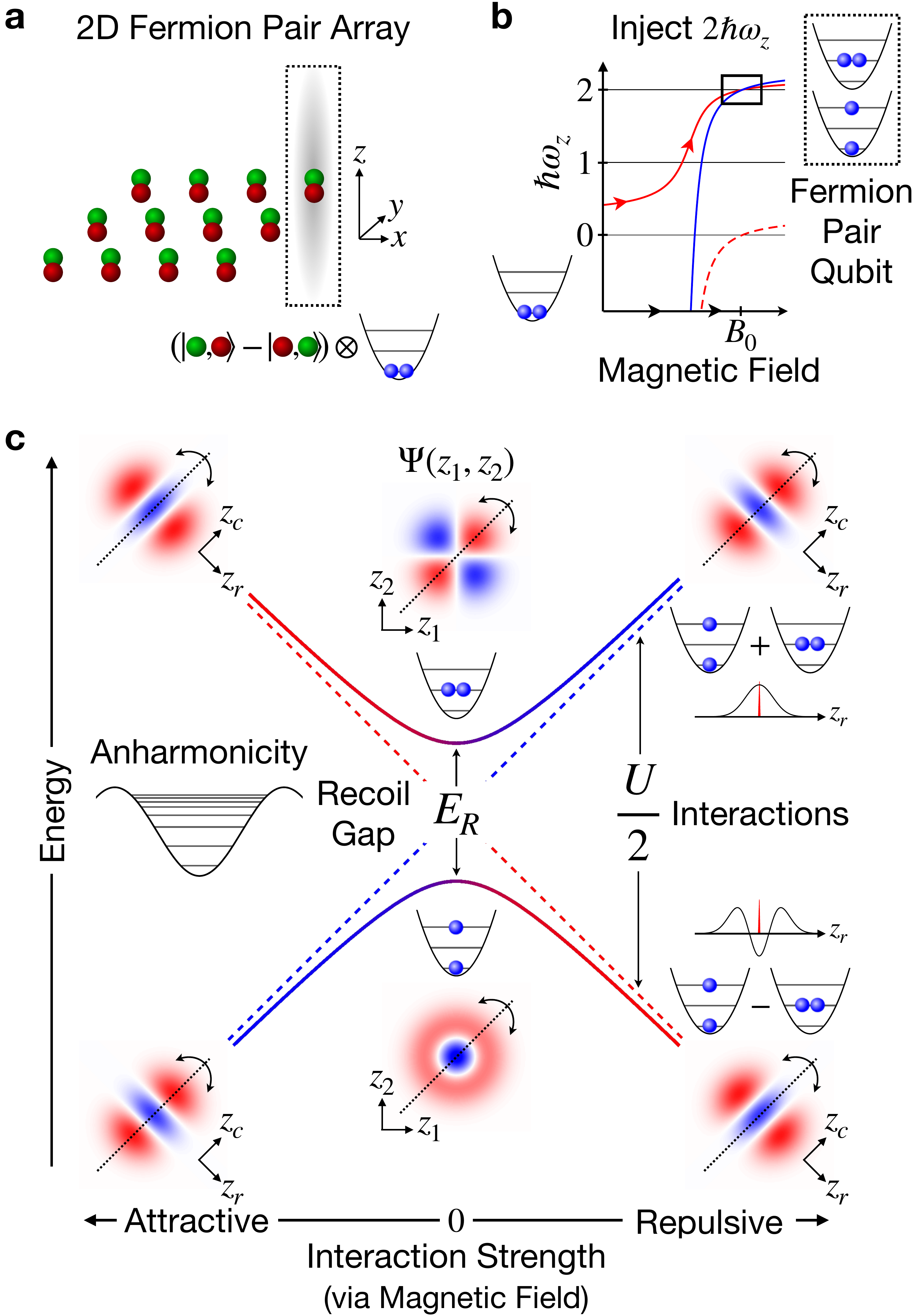}
	% \internallinenumbers
	\caption{\textbf{Spatial qubit encoding in a pair of entangled fermions.} 
	(a) Array of fermion pairs prepared from a two-component, fermionic quantum gas in an optical lattice. Each lattice site~(dashed box) initially contains two fermionic atoms in the spatial ground state of a 3D anisotropic harmonic trap, forming a spin singlet. Information is stored in the spatial wavefunction in the $z$ direction, which must remain exchange-symmetric $\Psi(z_1, z_2){=}\Psi(z_2, z_1)$ (blue spheres).
	(b) Injection of $2\hbar \omega_z$ vibrational energy via a magnetic field sweep across a Feshbach resonance brings the fermion pair~(red solid line) to the qubit subspace. For vanishing interatomic interactions, at field $B_0$, there are two degenerate two-particle states of harmonic motion along $z$, with atoms either both in the first excited state $\ket{1,1}$ or with one atom carrying two excitations $\ket{0,2}_{\rm s}$~(dashed box). 
	(c) Full control over the fermion pair qubit via trap anharmonicity and tunable interatomic interactions. At vanishing interactions (center), anharmonicity non-linearly reduces the energy of excited harmonic states (left schematic). States $\ket{1,1}$ (upper box) and $\ket{0,2}_{\rm s}$ (lower box) are split by the recoil energy $E_R {=} \hbar^2\pi^2/(2 m a_z^2)$, determined solely by geometry (lattice spacing $a_z$) and the atomic mass $m$. Strong interactions ($|U| {\gg} E_R$, left and right side) fully mix the anharmonic eigenstates. The fermion pair stores the $2\hbar\omega_z$ vibrational energy in either center-of-mass motion $z_c{=}(z_1{+}z_2)/\sqrt{2}$ (upper right box) or relative motion $z_r{=}(z_1{-}z_2)/\sqrt{2}$ (lower right box). This avoided crossing enables universal control of fermion pair qubits.
	}
	\label{fig:Fig_Schematic}
\end{figure}

To initialize the quantum register, we cool a two-state mixture of fermions into the lowest band of a 2D optical lattice. Increasing the lattice potential creates a large array of hundreds of fermion pairs in isolated wells, with each fermion occupying the 3D motional ground state of its well~(Fig.~\ref{fig:Fig_Schematic}(a)). Pauli exclusion is crucial in this preparation step: It energetically freezes out triply occupied sites and it forces the spin wavefunction of each fermion pair to be a spin singlet, thereby protecting the two-particle spatial wavefunction $\Psi(\vec{r}_1,\vec{r}_2)$ to remain exchange-symmetric at all times, $\Psi(\vec{r}_1,\vec{r}_2) = \Psi(\vec{r}_2,\vec{r}_1)$. Motion in each well occurs in the quasi-1D regime, with in-plane ($x$-$y$) confinement much stronger than in the out-of-plane ($z$) direction (angular frequencies $\omega_{x,y} \approx 4 \omega_z$). The quantum information is encoded in the subspace of vibrational states with two units of harmonic energy along $z$~($2\hbar\omega_z$), where exactly two symmetric two-particle states exist. In the first state $\ket{1,1}$ each atom carries one excitation, while in the second state $\ket{0,2}_{\rm s}{=}(\ket{0,2}{+}\ket{2,0})/\sqrt{2}$ one atom carries both quanta, leaving the other in the ground state. These two states differ only in how vibration is distributed within the atom pair, and are thus immune to fluctuations in $\hbar\omega_z$. Generally, single-particle perturbations can only couple the pair states at second order. The singlet spin wavefunction being a bystander to the pair motion, pairs are also immune against magnetic field noise.

To access this symmetry-protected subspace, we make use of a Feshbach resonance~\cite{Chin2010Feshbach} caused by an avoided crossing of the atom pair with a molecular state~\cite{Regal2003Creation}~(Fig.~\ref{fig:Fig_Schematic}(b)). Sweeping the magnetic field across the resonance injects $2\hbar\omega_z$ energy into the relative motion of the fermion pair~\cite{Kohl2005,Diener2006, Zurn2012} promoting it to state $\ket{2}_{\rm rel}$, while leaving the center-of-mass (COM) motion in the harmonic ground state, $\ket{0}_{\rm COM}$. At the magnetic field $B_0$ where interactions vanish, this atom pair in relative motion becomes degenerate with the state $\ket{2}_{\rm COM}\ket{0}_{\rm rel}$ where the two atoms instead have two units of COM motion~\cite{Busch1998, Idziaszek2005}. The latter emerges from a molecular state with that same excited COM motion existing at $B < B_0$.

Control of the fermion pair qubit is achieved via tunable interactions and the anharmonicity of the lattice potential, which couples the interacting states~\cite{Bolda2005,Sala2013, Sala2016Theory, Ishmukhamedov2017Tunneling}. Fig.~\ref{fig:Fig_Schematic}(c) shows a schematic of the energy spectrum and two-particle spatial wavefunctions $\Psi(z_1,z_2)$ of the fermion pair qubit states as interactions are tuned from attractive to vanishing to repulsive using the magnetic field. For each fermion pair state, $\Psi(z_1,z_2)$ is reflection symmetric about the line $z_1{=}z_2$.

\begin{figure*}[!t]
	\centering
	\includegraphics[width=\textwidth]{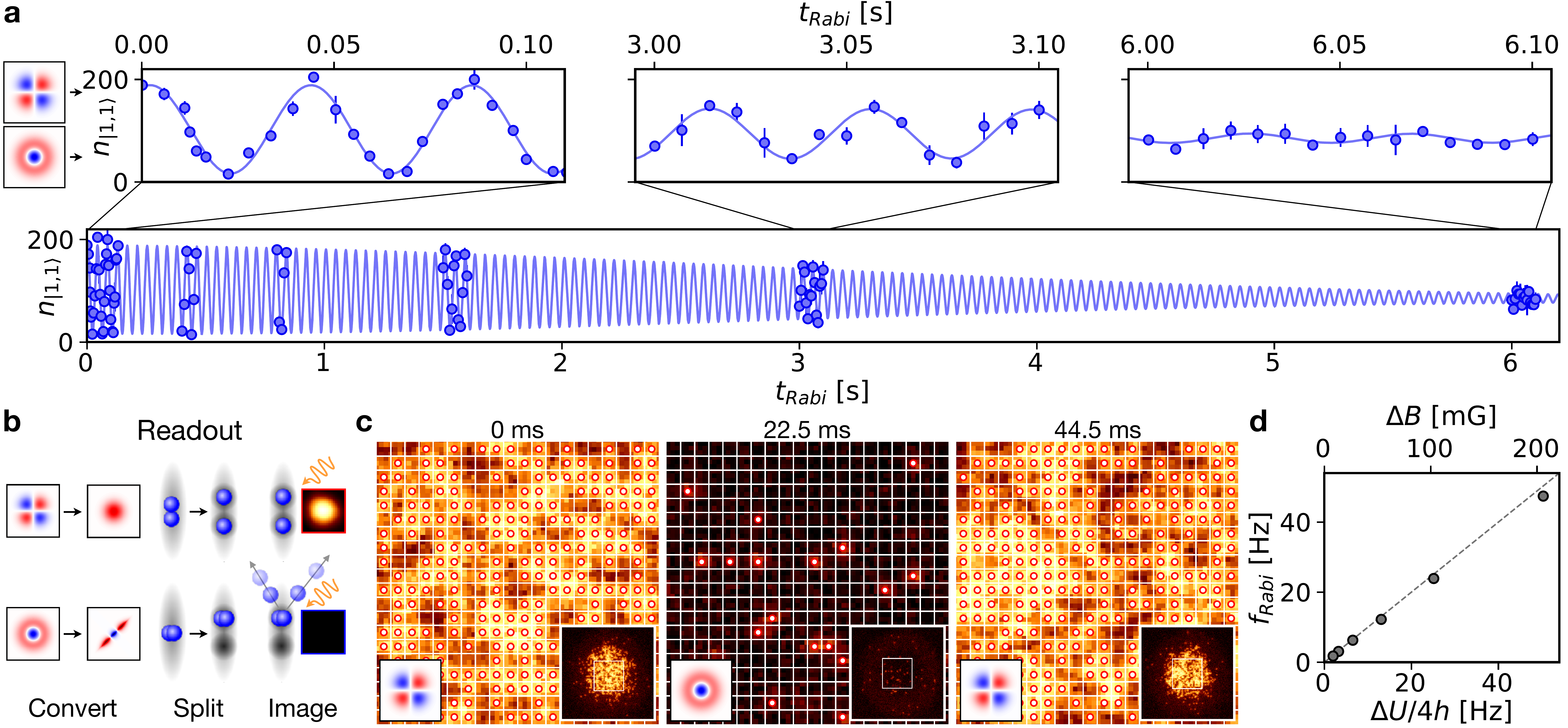}
	% \internallinenumbers
	\caption{\textbf{Simultaneous coherent manipulation and parallel readout of hundreds of motional fermion pair qubits.} 
	(a)~Modulation of interactions at the recoil gap drives a Rabi oscillation between fermion pair states $\ket{1,1}$ and $\ket{0,2}_{\rm s}$, measured via the number $n_{\ket{1,1}}$ of pairs in state $\ket{1,1}$ in a central region of the 2D array~(see methods).
	(b)~Readout proceeds by converting each fermion pair to either a molecule~(lower panel) or repulsive pair~(upper panel) using the Feshbach resonance. Application of a superlattice then splits repulsive pairs for fluorescence imaging, while tightly-bound molecules are ejected and appear dark~\cite{Hartke2020}.
	(c)~A single fluorescence image reveals each of the fermion pair qubits in parallel, with single-site resolution. The field of view displayed is $20\times20$ lattice sites, with lower right insets showing the entire atomic cloud.
	(d)~Measured Rabi frequency $f_{\rm Rabi}$ vs.~the measured magnetic field modulation $\Delta B$, and corresponding calculated interaction energy modulation $\Delta U/4$. The dashed line shows the predicted Rabi frequency $f_{\rm Rabi}{=}\Delta U/4h$, in good agreement with experiment. Here, and elsewhere, error bars on $n_{\ket{1,1}}$ show standard deviation from 2-3 repetitions. Error bars representing the fit error for $f_{\rm Rabi}$ are smaller than datapoints.
	}
	\label{fig:Fig_Rabi}
\end{figure*}

At vanishing interactions, the degeneracy of the two pair qubit states is lifted by the anharmonicity of the lattice~(center of Fig.~\ref{fig:Fig_Schematic}(c)). Each atom experiences an identical periodic potential $V E_R \sin(\pi z /a_z)^2$, where the lattice depth $V$ is given in units of the recoil energy $E_R {=} \hbar^2\pi^2/(2 m a_z^2)$, and $a_z$ is the lattice periodicity. For $V{\gg}1$ (in the experiment, $V{\approx}8000$), atoms experience a deep lattice and are localized near the potential minima. To first order, the trap is harmonic with energy $\hbar\omega_z{=}2E_R \sqrt{V}$, and the pair states $\ket{1,1}$ and $\ket{0,2}_{\rm s}$ are degenerate, removing dependence of their energy difference on $V$. At second order, the quartic corrections ${\sim }V E_R(z /a_z)^4$ possess two crucial effects. First, the characteristic size of the wave packet scales with the harmonic oscillator length $\sqrt{\hbar/m\omega_z}{=}(a_z /\pi) V^{-1/4}$, leading to quartic corrections ${\sim}E_R$ on the scale of the recoil energy, independent of the potential depth. Second,  because $\langle n| {z^4}|n\rangle$ grows as ${\sim}n^2$, the single-particle energy spectrum becomes non-linear $E_n{-}E_0{\approx}n\hbar\omega_z{-}n(n{+}1)E_R/2$, which breaks the degeneracy of the atom pair states. As a result, $\ket{0,2}_{\rm s}$ shifts below $\ket{1,1}$ by $E_R$, which we call the recoil gap. The relative fluctuations of this gap with $V$, due to the next order~\cite{SI} corrections $(9/8)E_R/\sqrt{V}$, are strongly suppressed by $\sqrt{V}{\gg}1$, highlighting the advantages of atom pair states compared to encoding information in single-particle harmonic states~\cite{Muller2007, Forster2009, VanFrank2014}. Moreover, the existence and inherent stability of the recoil gap are not specific to a lattice potential, and are a general feature of any anharmonic potential with rigid shape, with the role of $a_z$ replaced by another geometric scale.

When repulsive interactions dominate~(right side of Fig.~\ref{fig:Fig_Schematic}(c)), the fermion pair behaves like two pendula coupled by a spring, and the atoms oscillate with two quanta in either common or relative motion. Indeed, the higher energy pair state, when viewed along the center-of-mass and relative axes, is simply $|2\rangle_{\rm COM}|0\rangle_{\rm rel}{=} (\ket{0,2}_{\rm s}{+}\ket{1,1})/\sqrt{2}$, while the lower energy state is $|0\rangle_{\rm COM}|2\rangle_{\rm rel}{=}(\ket{0,2}_{\rm s}{-}\ket{1,1})/\sqrt{2}$. In the relative ground state $|0\rangle_{\rm rel}$ atoms overlap maximally and experience a repulsive energy shift~$\!\!\phantom{|}_{\rm rel}\langle 0|\hat{U}|0\rangle \phantom{|}_{\!\!\rm rel}{=}U$ from two-particle interactions $\hat{U}$, while in $|2\rangle_{\rm rel}$ they less strongly overlap, leading to a weaker shift~\cite{SI}~$\!\!\phantom{|}_{\rm rel}\langle 2|\hat{U}|2\rangle \phantom{|}_{\!\!\rm rel}{=}U/2$. The resulting energy separation between the pair states is $U/2$, which is set by the interaction strength and is experimentally controlled via the magnetic field.

In Fig.~\ref{fig:Fig_Rabi}(a) we demonstrate universal control of the fermion pair qubit by driving a Rabi oscillation between the recoil gap eigenstates $\ket{1,1}$ and $\ket{0,2}_{\rm s}$. Analogous to any two level system, Rabi oscillations are produced by modulating an off-diagonal matrix element, the interaction energy $\bra{1,1}\hat{U}\ket{0,2}_{\rm s} {=} U/4$, at the frequency of the recoil gap~\cite{Thompson2005Ultracold,SI}.
The oscillation is observed by counting the number $n_{\ket{1,1}}$ of fermion pairs in state $\ket{1,1}$ in a central region of the 2D array~(see methods). 
To detect the state of each pair qubit, we engineer state $\ket{1,1}$ to fluoresce when imaged, and state $\ket{0,2}_{\rm s}$ to appear dark~(Fig.~\ref{fig:Fig_Rabi}(b)). This is achieved by coherently splitting exclusively fermion pairs in $\ket{1,1}$ via a double-well into two separate, spin-entangled fermions, which are subsequently imaged.
With a single image of separated atom pairs under a quantum gas microscope~\cite{Hartke2020}, parallel readout of all fermion pair qubits in the 2D array is achieved with single-site resolution. Fig.~\ref{fig:Fig_Rabi}(c) shows the first complete Rabi cycle, as the 2D register array Rabi oscillates from $\ket{1,1}$~(bright) to $\ket{0,2}_{\rm s}$~(dark) and returns to $\ket{1,1}$. 

The measured Rabi frequency $f_{\rm Rabi}$ of the interaction-driven pair qubit agrees well, for moderate driving, with the interaction modulation $\Delta U/4$, calculated from the calibrated trap parameters and scattering properties~(see methods)~(Fig.~\ref{fig:Fig_Rabi}(d)). The robust energy separation between the pair qubit states and other symmetry-protected states of the fermion pair allows increasing the Rabi coupling to values at and beyond the recoil gap. In this non-perturbative regime of strong driving, the fermion pair exhibits a non-sinusoidal response~\cite{SI} which can be used for quantum control at rates exceeding the energy gap~\cite{Fuchs2009Gigahertz}.

\begin{figure}[!t]
	\centering
	\includegraphics[width=\columnwidth]{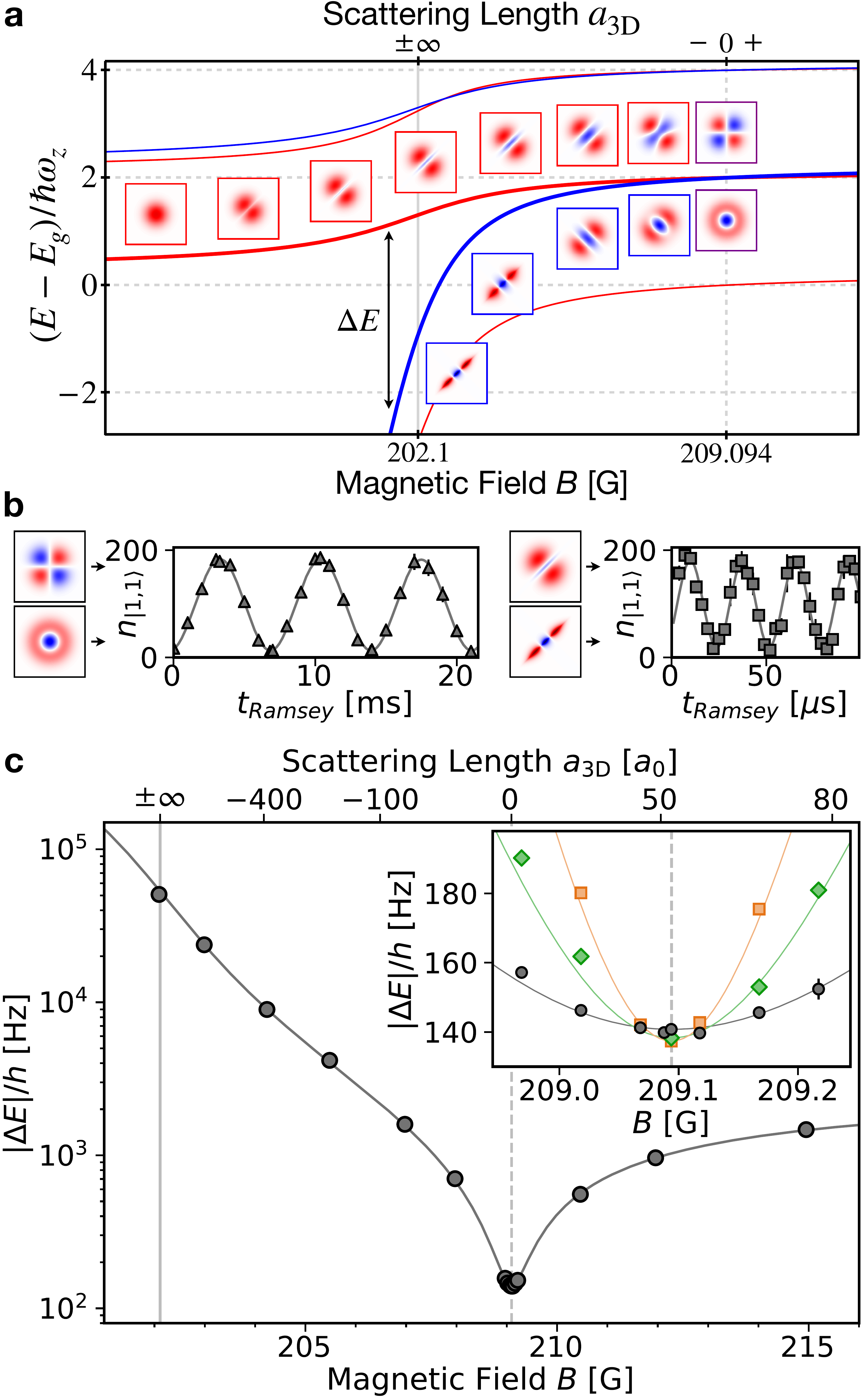}
	\vskip 2pt % removes extra space below figure
	% \internallinenumbers
	\caption{{\bf Crossover from fermion pair to molecule qubit.}
	(a)~Theoretical energy spectrum of an interacting fermion pair in the crossover to a molecular state. Pairs in the ground~(second excited) COM state are shown in red~(blue), with the qubit states highlighted (thick lines). Insets show approximate 1D pair qubit wavefunctions $\Psi(z_1,z_2)$ with a small trap anharmonicity. Energies $E$ are calculated for an anisotropic 3D harmonic trap with $\omega_x{=}\omega_y{=}3.853\,\omega_z$~\cite{Idziaszek2005}~(see methods). $E_g$ is the energy of the non-interacting ground state. 
	(b)~Ramsey interferometry measures the fermion pair qubit energy splitting, ranging from $|\Delta E|{=}h{\times}140.76(3)\,$Hz at vanishing interactions (left panel) to $h{\times}36.00(3)\,$kHz near the strongest explored interactions~(right panel,~$B{\approx}202.5\,$G).
	(c)~Measured energy splitting versus magnetic field, from vanishing to strongly attractive and repulsive interactions. The solid line is the theoretical prediction, using the measured recoil gap as an input, without fit parameters. The inset shows the energy splitting at the recoil gap, demonstrating insensitivity to doubling the harmonic frequency from $\omega_z/2\pi {=} 25.09(4)\,$kHz (black) and 38.50(1)$\,$kHz (green) to 51.8(1)$\,$kHz (orange). Error bars for $|\Delta E|$ represent sinusoidal fitting error, and are smaller than datapoints in the main figure.
	}
	\label{fig:Fig_Ramsey}
\end{figure}

We now demonstrate that the fermion pair qubit can coherently crossover into the molecular regime~\cite{Chin2010Feshbach} of tightly bound fermion pairs. This enables wide tunability of the pair qubit frequency and opens up applications for molecular clocks~\cite{Schiller2014Simplest,Kondov2019Molecular} and molecule-based quantum information protocols~\cite{DeMille2002Quantum}. Fig.~\ref{fig:Fig_Ramsey}(a) shows the calculated energy spectrum of the fermion pair in the molecular crossover~\cite{Busch1998, Idziaszek2005}. For every state of COM motion, there is an identical ladder of states of relative motion, starting with the molecular state~\cite{Sala2016Theory}. The pair qubit states approach degeneracy for vanishing interactions, at the zero of the scattering length $a_{\rm 3D}$~\cite{Ishmukhamedov2017Tunneling}. There, the approximate 1D two-particle spatial wavefunctions~(insets) approach the eigenstates $\ket{1,1}$ and $\ket{0,2}_{\rm s}$ of the recoil gap. For increasing attraction, moving towards the Feshbach resonance, the lower energy state first evolves into $|2\rangle_{\rm COM}|0\rangle_{\rm rel}$ and then into a deeply bound molecular state with COM motion, with the two atoms therefore strongly bound to each other, as seen by their wavefunction being concentrated near the diagonal $z_1{\approx}z_2$. In stark contrast, the higher energy state evolves into $|0\rangle_{\rm COM}|2\rangle_{\rm rel}$ and then into a strongly repulsive, ``fermionized'' pair~\cite{Kohl2005, Zurn2012}, with largely reduced probability for the two atoms to be at the same location. Near the point of fermionization, additional, narrow anharmonicity-mediated resonances exist with molecules in excited transverse COM states~\cite{Haller2010}, where coherent interconversion has been demonstrated~\cite{Sala2013}.

To probe the energy spectrum of the fermion pair qubit, we perform Ramsey interferometry of the register states~(Fig.~\ref{fig:Fig_Ramsey}(b)). At vanishing interactions (left panel) anharmonicity interferes the pair states $\ket{1,1}$ and $\ket{0,2}_{\rm s}$ at the recoil gap frequency $E_R/h{=}140.76(3)\,$Hz. In contrast, at strong interactions (right panel) the molecular binding energy drives a much faster Ramsey oscillation at 36.00(3)$\,$kHz between a 2D array of strongly repulsive fermion pairs and a lattice of tightly-bound molecules~\cite{Donley2002,Thompson2005Ultracold,Syassen2007}. Ramsey measurements across the Feshbach resonance~(Fig.~\ref{fig:Fig_Ramsey}(c)) demonstrate a dramatic ability to tune the frequency of the fermion pair qubit, and thus the entangling speed of the motion of fermion pairs, over multiple orders of magnitude.

A key feature of the recoil gap is suppressed sensitivity to laser intensity noise. The inset of Fig.~\ref{fig:Fig_Ramsey}(c) shows the energy gap near zero interactions at different lattice depths. As the trap depth is increased four-fold, the harmonic energy $\hbar\omega_z$ doubles, while the recoil gap energy changes by only $2.40(6)\%$. Moving away from the avoided crossing, the energy difference begins to be determined by interactions. In this regime, increasing the trapping depth confines the atomic wavefunction more tightly and enhances interactions, thus enabling local manipulation of the pair qubit frequency with targeted laser beams~\cite{Weitenberg2011Single}. 

\begin{figure*}[t]
	\centering
	\includegraphics[width=\textwidth]{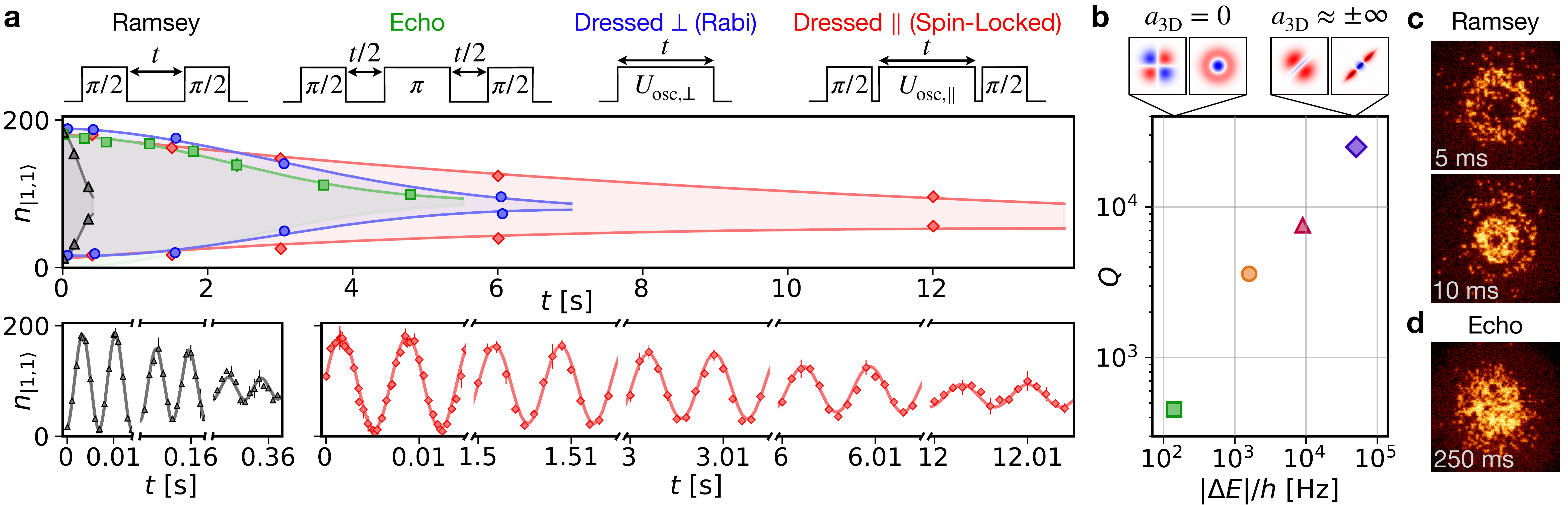}
	% \internallinenumbers
	\caption{{\bf Second-scale coherence of the fermion pair qubit.}
	(a)~A Ramsey oscillation at the recoil gap has a Gaussian envelope (black triangles) with $1/e$~timescale~$\tau{=}300(10)\,$ms (time series shown below). Inserting an echo $\pi$-pulse in a Ramsey oscillation~(green~squares) extends coherence to~$\tau_{\rm Echo}{=}3.2(1)\,$s (Gaussian fit), indicating that phase noise is static. Dressing the pair qubit at the recoil gap with modulated interactions suppresses static noise, extending the coherence time to $\tau{=}4.0(3)\,$s for a state which is prepared perpendicular to the drive~(Rabi oscillation, blue~circles, Gaussian fit) and to $\tau {=} 8.5(5)\,$s for a state which is prepared parallel to the drive~(spin-locked oscillation, red~diamonds, exponential fit). $U_{\rm osc_\perp}$ denotes a Rabi drive with coupling $f_{\rm Rabi}{=}23.902(4)\,{\rm Hz}$, while $U_{\rm osc,\parallel}$ denotes the same drive preceded by a quarter period of free evolution~\cite{SI}.
	(b)~The number of coherent oscillations of the fermion pair qubit per decay time, given by the quality factor $Q {=} |\Delta E|{\,}\tau_{\rm Echo}/h$, grows in the molecular crossover of the fermion pair, from $Q{\approx}450$ at vanishing interactions (green~square) to $Q{\approx}25000$ at $B{=}202.091(8)\,$G, near the Feshbach resonance (purple~diamond,~$\tau_{\rm Echo}{=}0.49(2)\,$s)~\cite{SI}.
	(c)~Small interaction energy variations due to the curvature of the lattice beams produce rings in a Ramsey oscillation at $B{=}202.091(8)\,$G after $10{\,}$ms~(${\sim}500$ oscillations). (d) A 250${\,}$ms echo sequence (${\sim}12500$~oscillations) unwinds these rings to recover spatial coherence. Envelope data in 
	(a), except for echo data, are the extreme values of an independent sinusoidal fit over several cycles at each time. Echo data directly obtains the contrast. Shaded regions and decay times $\tau$ are extracted from simultaneous fits to all data. Error bars from fit error in (a) and (b) are generally smaller than datapoints. 
	}
	\label{fig:Fig_Coherence}
\end{figure*}

Finally, we explore the coherence of the fermion pair qubit, beginning in Fig.~\ref{fig:Fig_Coherence}(a) with the properties of the recoil gap states. The decay envelope of various experimental sequences~(upper schematics) provides insight into existing noise sources, guiding methods to improve coherence. As a first measurement, we perform a Ramsey experiment which detects intrinsic phase noise. The superposition decays in $\tau{=}300(10)\,$ms after ${\sim}40$ recoil oscillations, with random spatial structure. In contrast, if an echo $\pi$-pulse is inserted into the Ramsey oscillation to cancel static noise, coherence is extended to $\tau_{\rm Echo}{=}3.2(1)\,$s, corresponding to ${\sim}450$ recoil oscillations and $10^5$ harmonic oscillator periods. In combination, these observations indicate the presence of a near static spatial variation of recoil gap energies on the order of ${\sim} 0.75\,$Hz, or $0.5\%\,E_R$, within the 2D array. 

We eliminate the effects of static noise at the recoil gap by modulating interactions at the recoil gap frequency. This dressing scheme preserves the quantum information in an arbitrary qubit state for over 4 seconds. The method is analogous to applying an oscillating transverse magnetic field in order to stabilize an ensemble of dephasing spins~\cite{Hartmann1962Nuclear}. In the frame rotating with the drive, the applied static field is perpendicular to and much larger than any residual dephasing fields~\cite{Hartmann1962Nuclear,Rabl2009Strong}, leading to a uniform quantization energy across the ensemble, thereby preserving arbitrary state superpositions. Here, interaction dressing extends coherence to $\tau{=}4.0(3)\,$s for a state prepared perpendicular to the drive, corresponding to a Rabi oscillation. If the pair qubit is instead first rotated to align with the drive, in spin-locked operation, a further extension of coherence to $\tau{=}8.5(5)\,$s is observed, and oscillations are still visible after $12$ seconds with good signal to noise. As expected for this driven operation~\cite{Hartmann1962Nuclear}, which increases the coherence from the dephasing time $T_2$ to the population decay time $T_1$, these decoherence rates approach the natural limits provided by the measured bit-flip rates and loss rates for the two pair qubit states~(see methods).

To characterize coherence of the fermion pair qubit in the molecular crossover, we measure the quality factor $Q {=}|\Delta E|{\,}\tau_{\rm Echo}/h$ as a function of interaction strength and thus of the fermion pair qubit energy splitting $|\Delta E|$~(Fig.~\ref{fig:Fig_Coherence}(b)). Remarkably, near the scattering resonance a superposition of a repulsive fermion pair and a tightly-bound molecule remains coherent for ${\sim}25000$ Ramsey oscillations, a promising value for quantum error correction based on two-fermion qubits~\cite{DiVincenzo2000Criteria}. The dominant source of decoherence is the underlying curvature of the optical lattice beams, which leads to a radial variation of the interaction energy of ${\sim}0.2\%$ at 10 sites from the center of the array. This can be observed as spatial rings in a Ramsey experiment near the Feshbach resonance, as in Fig.~\ref{fig:Fig_Coherence}(c), taken after ${\sim}500$ oscillations. With an echo pulse~(Fig.~\ref{fig:Fig_Coherence}(d)), these rings refocus spatially even after ${\sim}12500$ oscillations.

In summary, we have demonstrated a quantum register based on fermion pairs in a 2D optical lattice. Our method provides a new route towards quantum computation and simulation by leveraging Pauli exclusion for high fidelity preparation and control of spin- and motionally entangled states of fermions.
Allowing for tunneling between adjacent sites enables the exploration of extended Fermi-Hubbard models with additional orbital degrees of freedom~\cite{Ho2006,Koga2004Orbital,Kubo2007Pairing}.
Full gate-based control of entangled many-body states may be realized via cold collisions~\cite{Mandel2003controlled,Anderlini2007,Greif2013Short,Mamaev2020Quantum} or Rydberg excitation~\cite{Labuhn2016Tunable,Bernien2017Probing,Hollerith2019}. Furthermore, the accurate simultaneous control of hundreds of molecules in superposition with free atom pairs allows for site-resolved detection of many-body states of dipolar molecules~\cite{Yan2013Observation}, for tests of fundamental symmetries~\cite{Baron2014Order, Cairncross2017Precision}, and metrology based on molecular clocks~\cite{Schiller2014Simplest,Kondov2019Molecular}.

% {\bf Acknowledgements:}
We would like to thank Carsten Robens for stimulating discussions. This work was supported by the NSF through the Center for Ultracold Atoms and Grant PHY-2012110, ONR (Grant No. N00014-17-1-2257), AFOSR (Grant No. FA9550-16-1-0324), AFOSR-MURIs on Quantum Phases of Matter (Grant
No. FA9550-14-1-0035) and on Molecular Ensembles, the Gordon and Betty Moore Foundation through grant GBMF5279, and the Vannevar Bush Faculty Fellowship. M.Z. acknowledges support from the Alexander von Humboldt Foundation.

% {\bf Author contributions:}
% The experiment was designed by all authors. T.H., B.O., and N.J.~collected and analyzed the data. All authors contributed to the manuscript.

% {\bf Author information:}
% The authors declare no competing financial interests. 
Correspondence and requests for materials should be addressed to T.H.~(hartke@mit.edu) and M.Z.~(zwierlein@mit.edu).

\bibliography{FermionicQuantumRegister}

\begin{thebibliography}{78}%
\makeatletter
\providecommand \@ifxundefined [1]{%
 \@ifx{#1\undefined}
}%
\providecommand \@ifnum [1]{%
 \ifnum #1\expandafter \@firstoftwo
 \else \expandafter \@secondoftwo
 \fi
}%
\providecommand \@ifx [1]{%
 \ifx #1\expandafter \@firstoftwo
 \else \expandafter \@secondoftwo
 \fi
}%
\providecommand \natexlab [1]{#1}%
\providecommand \enquote  [1]{``#1''}%
\providecommand \bibnamefont  [1]{#1}%
\providecommand \bibfnamefont [1]{#1}%
\providecommand \citenamefont [1]{#1}%
\providecommand \href@noop [0]{\@secondoftwo}%
\providecommand \href [0]{\begingroup \@sanitize@url \@href}%
\providecommand \@href[1]{\@@startlink{#1}\@@href}%
\providecommand \@@href[1]{\endgroup#1\@@endlink}%
\providecommand \@sanitize@url [0]{\catcode `\\12\catcode `\$12\catcode
  `\&12\catcode `\#12\catcode `\^12\catcode `\_12\catcode `\%12\relax}%
\providecommand \@@startlink[1]{}%
\providecommand \@@endlink[0]{}%
\providecommand \url  [0]{\begingroup\@sanitize@url \@url }%
\providecommand \@url [1]{\endgroup\@href {#1}{\urlprefix }}%
\providecommand \urlprefix  [0]{URL }%
\providecommand \Eprint [0]{\href }%
\providecommand \doibase [0]{http://dx.doi.org/}%
\providecommand \selectlanguage [0]{\@gobble}%
\providecommand \bibinfo  [0]{\@secondoftwo}%
\providecommand \bibfield  [0]{\@secondoftwo}%
\providecommand \translation [1]{[#1]}%
\providecommand \BibitemOpen [0]{}%
\providecommand \bibitemStop [0]{}%
\providecommand \bibitemNoStop [0]{.\EOS\space}%
\providecommand \EOS [0]{\spacefactor3000\relax}%
\providecommand \BibitemShut  [1]{\csname bibitem#1\endcsname}%
\let\auto@bib@innerbib\@empty
%</preamble>
\bibitem [{\citenamefont {Inguscio}\ \emph {et~al.}(2008)\citenamefont
  {Inguscio}, \citenamefont {Ketterle},\ and\ \citenamefont
  {Salomon}}]{Inguscio2008}%
  \BibitemOpen
  \bibinfo {editor} {\bibfnamefont {M.}~\bibnamefont {Inguscio}}, \bibinfo
  {editor} {\bibfnamefont {W.}~\bibnamefont {Ketterle}}, \ and\ \bibinfo
  {editor} {\bibfnamefont {C.}~\bibnamefont {Salomon}},\ eds.,\ \href@noop {}
  {\emph {\bibinfo {title} {{Ultracold Fermi Gases}}}}\ (\bibinfo  {publisher}
  {IOS Press},\ \bibinfo {address} {Amsterdam},\ \bibinfo {year}
  {2008})\BibitemShut {NoStop}%
\bibitem [{\citenamefont {Bloch}\ \emph {et~al.}(2008)\citenamefont {Bloch},
  \citenamefont {Dalibard},\ and\ \citenamefont {Zwerger}}]{Bloch2008Many}%
  \BibitemOpen
  \bibfield  {author} {\bibinfo {author} {\bibfnamefont {I.}~\bibnamefont
  {Bloch}}, \bibinfo {author} {\bibfnamefont {J.}~\bibnamefont {Dalibard}}, \
  and\ \bibinfo {author} {\bibfnamefont {W.}~\bibnamefont {Zwerger}},\ }\href
  {\doibase 10.1103/RevModPhys.80.885} {\bibfield  {journal} {\bibinfo
  {journal} {Rev. Mod. Phys.}\ }\textbf {\bibinfo {volume} {80}},\ \bibinfo
  {pages} {885} (\bibinfo {year} {2008})}\BibitemShut {NoStop}%
\bibitem [{\citenamefont {Zwerger}(2012)}]{Zwerger2012}%
  \BibitemOpen
  \bibinfo {editor} {\bibfnamefont {W.}~\bibnamefont {Zwerger}},\ ed.,\
  \href@noop {} {\emph {\bibinfo {title} {The BCS-BEC Crossover and the Unitary
  Fermi Gas}}}\ (\bibinfo  {publisher} {Springer Berlin Heidelberg},\ \bibinfo
  {year} {2012})\BibitemShut {NoStop}%
\bibitem [{\citenamefont {Inguscio}\ \emph {et~al.}(2016)\citenamefont
  {Inguscio}, \citenamefont {Ketterle}, \citenamefont {Stringari},\ and\
  \citenamefont {Roati}}]{Inguscio2016}%
  \BibitemOpen
  \bibinfo {editor} {\bibfnamefont {M.}~\bibnamefont {Inguscio}}, \bibinfo
  {editor} {\bibfnamefont {W.}~\bibnamefont {Ketterle}}, \bibinfo {editor}
  {\bibfnamefont {S.}~\bibnamefont {Stringari}}, \ and\ \bibinfo {editor}
  {\bibfnamefont {G.}~\bibnamefont {Roati}},\ eds.,\ \href@noop {} {\emph
  {\bibinfo {title} {{Quantum matter at ultralow temperatures}}}}\ (\bibinfo
  {publisher} {IOS Press},\ \bibinfo {address} {Amsterdam},\ \bibinfo {year}
  {2016})\BibitemShut {NoStop}%
\bibitem [{\citenamefont {Gross}\ and\ \citenamefont
  {Bloch}(2017)}]{Gross2017QSim}%
  \BibitemOpen
  \bibfield  {author} {\bibinfo {author} {\bibfnamefont {C.}~\bibnamefont
  {Gross}}\ and\ \bibinfo {author} {\bibfnamefont {I.}~\bibnamefont {Bloch}},\
  }\href {\doibase 10.1126/science.aal3837} {\bibfield  {journal} {\bibinfo
  {journal} {Science}\ }\textbf {\bibinfo {volume} {357}},\ \bibinfo {pages}
  {995} (\bibinfo {year} {2017})}\BibitemShut {NoStop}%
\bibitem [{\citenamefont {Abrams}\ and\ \citenamefont
  {Lloyd}(1997)}]{Abrams1997Simulation}%
  \BibitemOpen
  \bibfield  {author} {\bibinfo {author} {\bibfnamefont {D.~S.}\ \bibnamefont
  {Abrams}}\ and\ \bibinfo {author} {\bibfnamefont {S.}~\bibnamefont {Lloyd}},\
  }\href {\doibase 10.1103/PhysRevLett.79.2586} {\bibfield  {journal} {\bibinfo
   {journal} {Phys. Rev. Lett.}\ }\textbf {\bibinfo {volume} {79}},\ \bibinfo
  {pages} {2586} (\bibinfo {year} {1997})}\BibitemShut {NoStop}%
\bibitem [{\citenamefont {Bravyi}\ and\ \citenamefont
  {Kitaev}(2002)}]{Bravyi2002Fermionic}%
  \BibitemOpen
  \bibfield  {author} {\bibinfo {author} {\bibfnamefont {S.~B.}\ \bibnamefont
  {Bravyi}}\ and\ \bibinfo {author} {\bibfnamefont {A.~Y.}\ \bibnamefont
  {Kitaev}},\ }\href {\doibase 10.1006/aphy.2002.6254} {\bibfield  {journal}
  {\bibinfo  {journal} {Ann. Phys. (N. Y).}\ }\textbf {\bibinfo {volume}
  {298}},\ \bibinfo {pages} {210} (\bibinfo {year} {2002})}\BibitemShut
  {NoStop}%
\bibitem [{\citenamefont {Bauer}\ \emph {et~al.}(2020)\citenamefont {Bauer},
  \citenamefont {Bravyi}, \citenamefont {Motta},\ and\ \citenamefont
  {Kin-Lic~Chan}}]{Bauer2020Quantum}%
  \BibitemOpen
  \bibfield  {author} {\bibinfo {author} {\bibfnamefont {B.}~\bibnamefont
  {Bauer}}, \bibinfo {author} {\bibfnamefont {S.}~\bibnamefont {Bravyi}},
  \bibinfo {author} {\bibfnamefont {M.}~\bibnamefont {Motta}}, \ and\ \bibinfo
  {author} {\bibfnamefont {G.}~\bibnamefont {Kin-Lic~Chan}},\ }\href {\doibase
  10.1021/acs.chemrev.9b00829} {\bibfield  {journal} {\bibinfo  {journal}
  {Chem. Rev.}\ }\textbf {\bibinfo {volume} {120}},\ \bibinfo {pages} {12685}
  (\bibinfo {year} {2020})}\BibitemShut {NoStop}%
\bibitem [{\citenamefont {Kitaev}(2001)}]{Kitaev2001Unpaired}%
  \BibitemOpen
  \bibfield  {author} {\bibinfo {author} {\bibfnamefont {A.~Y.}\ \bibnamefont
  {Kitaev}},\ }\href {\doibase 10.1070/1063-7869/44/10S/S29} {\bibfield
  {journal} {\bibinfo  {journal} {Phys.-Uspekhi}\ }\textbf {\bibinfo {volume}
  {44}},\ \bibinfo {pages} {131} (\bibinfo {year} {2001})}\BibitemShut
  {NoStop}%
\bibitem [{\citenamefont {Bakr}\ \emph {et~al.}(2009)\citenamefont {Bakr},
  \citenamefont {Gillen}, \citenamefont {Peng}, \citenamefont {F{\"{o}}lling},\
  and\ \citenamefont {Greiner}}]{Bakr2009}%
  \BibitemOpen
  \bibfield  {author} {\bibinfo {author} {\bibfnamefont {W.~S.}\ \bibnamefont
  {Bakr}}, \bibinfo {author} {\bibfnamefont {J.~I.}\ \bibnamefont {Gillen}},
  \bibinfo {author} {\bibfnamefont {A.}~\bibnamefont {Peng}}, \bibinfo {author}
  {\bibfnamefont {S.}~\bibnamefont {F{\"{o}}lling}}, \ and\ \bibinfo {author}
  {\bibfnamefont {M.}~\bibnamefont {Greiner}},\ }\href {\doibase
  10.1038/nature08482} {\bibfield  {journal} {\bibinfo  {journal} {Nature}\
  }\textbf {\bibinfo {volume} {462}},\ \bibinfo {pages} {74} (\bibinfo {year}
  {2009})}\BibitemShut {NoStop}%
\bibitem [{\citenamefont {Sherson}\ \emph {et~al.}(2010)\citenamefont
  {Sherson}, \citenamefont {Weitenberg}, \citenamefont {Endres}, \citenamefont
  {Cheneau}, \citenamefont {Bloch},\ and\ \citenamefont {Kuhr}}]{Sherson2010}%
  \BibitemOpen
  \bibfield  {author} {\bibinfo {author} {\bibfnamefont {J.~F.}\ \bibnamefont
  {Sherson}}, \bibinfo {author} {\bibfnamefont {C.}~\bibnamefont {Weitenberg}},
  \bibinfo {author} {\bibfnamefont {M.}~\bibnamefont {Endres}}, \bibinfo
  {author} {\bibfnamefont {M.}~\bibnamefont {Cheneau}}, \bibinfo {author}
  {\bibfnamefont {I.}~\bibnamefont {Bloch}}, \ and\ \bibinfo {author}
  {\bibfnamefont {S.}~\bibnamefont {Kuhr}},\ }\href {\doibase
  10.1038/nature09378} {\bibfield  {journal} {\bibinfo  {journal} {Nature}\
  }\textbf {\bibinfo {volume} {467}},\ \bibinfo {pages} {68} (\bibinfo {year}
  {2010})}\BibitemShut {NoStop}%
\bibitem [{\citenamefont {Cheuk}\ \emph {et~al.}(2015)\citenamefont {Cheuk},
  \citenamefont {Nichols}, \citenamefont {Okan}, \citenamefont {Gersdorf},
  \citenamefont {Ramasesh}, \citenamefont {Bakr}, \citenamefont {Lompe},\ and\
  \citenamefont {Zwierlein}}]{Cheuk2015Quantum}%
  \BibitemOpen
  \bibfield  {author} {\bibinfo {author} {\bibfnamefont {L.~W.}\ \bibnamefont
  {Cheuk}}, \bibinfo {author} {\bibfnamefont {M.~A.}\ \bibnamefont {Nichols}},
  \bibinfo {author} {\bibfnamefont {M.}~\bibnamefont {Okan}}, \bibinfo {author}
  {\bibfnamefont {T.}~\bibnamefont {Gersdorf}}, \bibinfo {author}
  {\bibfnamefont {V.~V.}\ \bibnamefont {Ramasesh}}, \bibinfo {author}
  {\bibfnamefont {W.~S.}\ \bibnamefont {Bakr}}, \bibinfo {author}
  {\bibfnamefont {T.}~\bibnamefont {Lompe}}, \ and\ \bibinfo {author}
  {\bibfnamefont {M.~W.}\ \bibnamefont {Zwierlein}},\ }\href {\doibase
  10.1103/PhysRevLett.114.193001} {\bibfield  {journal} {\bibinfo  {journal}
  {Phys. Rev. Lett.}\ }\textbf {\bibinfo {volume} {114}},\ \bibinfo {pages}
  {193001} (\bibinfo {year} {2015})}\BibitemShut {NoStop}%
\bibitem [{\citenamefont {Haller}\ \emph {et~al.}(2015)\citenamefont {Haller},
  \citenamefont {Hudson}, \citenamefont {Kelly}, \citenamefont {Cotta},
  \citenamefont {Peaudecerf}, \citenamefont {Bruce},\ and\ \citenamefont
  {Kuhr}}]{Haller2015Single}%
  \BibitemOpen
  \bibfield  {author} {\bibinfo {author} {\bibfnamefont {E.}~\bibnamefont
  {Haller}}, \bibinfo {author} {\bibfnamefont {J.}~\bibnamefont {Hudson}},
  \bibinfo {author} {\bibfnamefont {A.}~\bibnamefont {Kelly}}, \bibinfo
  {author} {\bibfnamefont {D.~A.}\ \bibnamefont {Cotta}}, \bibinfo {author}
  {\bibfnamefont {B.}~\bibnamefont {Peaudecerf}}, \bibinfo {author}
  {\bibfnamefont {G.~D.}\ \bibnamefont {Bruce}}, \ and\ \bibinfo {author}
  {\bibfnamefont {S.}~\bibnamefont {Kuhr}},\ }\href {\doibase
  10.1038/nphys3403} {\bibfield  {journal} {\bibinfo  {journal} {Nat. Phys.}\
  }\textbf {\bibinfo {volume} {11}},\ \bibinfo {pages} {738} (\bibinfo {year}
  {2015})}\BibitemShut {NoStop}%
\bibitem [{\citenamefont {Parsons}\ \emph {et~al.}(2015)\citenamefont
  {Parsons}, \citenamefont {Huber}, \citenamefont {Mazurenko}, \citenamefont
  {Chiu}, \citenamefont {Setiawan}, \citenamefont {Wooley-Brown}, \citenamefont
  {Blatt},\ and\ \citenamefont {Greiner}}]{Parsons2015}%
  \BibitemOpen
  \bibfield  {author} {\bibinfo {author} {\bibfnamefont {M.~F.}\ \bibnamefont
  {Parsons}}, \bibinfo {author} {\bibfnamefont {F.}~\bibnamefont {Huber}},
  \bibinfo {author} {\bibfnamefont {A.}~\bibnamefont {Mazurenko}}, \bibinfo
  {author} {\bibfnamefont {C.~S.}\ \bibnamefont {Chiu}}, \bibinfo {author}
  {\bibfnamefont {W.}~\bibnamefont {Setiawan}}, \bibinfo {author}
  {\bibfnamefont {K.}~\bibnamefont {Wooley-Brown}}, \bibinfo {author}
  {\bibfnamefont {S.}~\bibnamefont {Blatt}}, \ and\ \bibinfo {author}
  {\bibfnamefont {M.}~\bibnamefont {Greiner}},\ }\href {\doibase
  10.1103/PhysRevLett.114.213002} {\bibfield  {journal} {\bibinfo  {journal}
  {Phys. Rev. Lett.}\ }\textbf {\bibinfo {volume} {114}},\ \bibinfo {pages}
  {213002} (\bibinfo {year} {2015})}\BibitemShut {NoStop}%
\bibitem [{\citenamefont {Omran}\ \emph {et~al.}(2015)\citenamefont {Omran},
  \citenamefont {Boll}, \citenamefont {Hilker}, \citenamefont {Kleinlein},
  \citenamefont {Salomon}, \citenamefont {Bloch},\ and\ \citenamefont
  {Gross}}]{Omran2015}%
  \BibitemOpen
  \bibfield  {author} {\bibinfo {author} {\bibfnamefont {A.}~\bibnamefont
  {Omran}}, \bibinfo {author} {\bibfnamefont {M.}~\bibnamefont {Boll}},
  \bibinfo {author} {\bibfnamefont {T.~A.}\ \bibnamefont {Hilker}}, \bibinfo
  {author} {\bibfnamefont {K.}~\bibnamefont {Kleinlein}}, \bibinfo {author}
  {\bibfnamefont {G.}~\bibnamefont {Salomon}}, \bibinfo {author} {\bibfnamefont
  {I.}~\bibnamefont {Bloch}}, \ and\ \bibinfo {author} {\bibfnamefont
  {C.}~\bibnamefont {Gross}},\ }\href {\doibase 10.1103/PhysRevLett.115.263001}
  {\bibfield  {journal} {\bibinfo  {journal} {Phys. Rev. Lett.}\ }\textbf
  {\bibinfo {volume} {115}},\ \bibinfo {pages} {263001} (\bibinfo {year}
  {2015})}\BibitemShut {NoStop}%
\bibitem [{\citenamefont {Edge}\ \emph {et~al.}(2015)\citenamefont {Edge},
  \citenamefont {Anderson}, \citenamefont {Jervis}, \citenamefont {McKay},
  \citenamefont {Day}, \citenamefont {Trotzky},\ and\ \citenamefont
  {Thywissen}}]{Edge2015}%
  \BibitemOpen
  \bibfield  {author} {\bibinfo {author} {\bibfnamefont {G.~J.~A.}\
  \bibnamefont {Edge}}, \bibinfo {author} {\bibfnamefont {R.}~\bibnamefont
  {Anderson}}, \bibinfo {author} {\bibfnamefont {D.}~\bibnamefont {Jervis}},
  \bibinfo {author} {\bibfnamefont {D.~C.}\ \bibnamefont {McKay}}, \bibinfo
  {author} {\bibfnamefont {R.}~\bibnamefont {Day}}, \bibinfo {author}
  {\bibfnamefont {S.}~\bibnamefont {Trotzky}}, \ and\ \bibinfo {author}
  {\bibfnamefont {J.~H.}\ \bibnamefont {Thywissen}},\ }\href {\doibase
  10.1103/PhysRevA.92.063406} {\bibfield  {journal} {\bibinfo  {journal} {Phys.
  Rev. A}\ }\textbf {\bibinfo {volume} {92}},\ \bibinfo {pages} {063406}
  (\bibinfo {year} {2015})}\BibitemShut {NoStop}%
\bibitem [{\citenamefont {Brown}\ \emph {et~al.}(2017)\citenamefont {Brown},
  \citenamefont {Mitra}, \citenamefont {Guardado-Sanchez}, \citenamefont
  {Schau{\ss}}, \citenamefont {Kondov}, \citenamefont {Khatami}, \citenamefont
  {Paiva}, \citenamefont {Trivedi}, \citenamefont {Huse},\ and\ \citenamefont
  {Bakr}}]{Brown2017Spin}%
  \BibitemOpen
  \bibfield  {author} {\bibinfo {author} {\bibfnamefont {P.~T.}\ \bibnamefont
  {Brown}}, \bibinfo {author} {\bibfnamefont {D.}~\bibnamefont {Mitra}},
  \bibinfo {author} {\bibfnamefont {E.}~\bibnamefont {Guardado-Sanchez}},
  \bibinfo {author} {\bibfnamefont {P.}~\bibnamefont {Schau{\ss}}}, \bibinfo
  {author} {\bibfnamefont {S.~S.}\ \bibnamefont {Kondov}}, \bibinfo {author}
  {\bibfnamefont {E.}~\bibnamefont {Khatami}}, \bibinfo {author} {\bibfnamefont
  {T.}~\bibnamefont {Paiva}}, \bibinfo {author} {\bibfnamefont
  {N.}~\bibnamefont {Trivedi}}, \bibinfo {author} {\bibfnamefont {D.~A.}\
  \bibnamefont {Huse}}, \ and\ \bibinfo {author} {\bibfnamefont {W.~S.}\
  \bibnamefont {Bakr}},\ }\href {\doibase 10.1126/science.aam7838} {\bibfield
  {journal} {\bibinfo  {journal} {Science}\ }\textbf {\bibinfo {volume}
  {357}},\ \bibinfo {pages} {1385} (\bibinfo {year} {2017})}\BibitemShut
  {NoStop}%
\bibitem [{\citenamefont {Koepsell}\ \emph {et~al.}(2020)\citenamefont
  {Koepsell}, \citenamefont {Hirthe}, \citenamefont {Bourgund}, \citenamefont
  {Sompet}, \citenamefont {Vijayan}, \citenamefont {Salomon}, \citenamefont
  {Gross},\ and\ \citenamefont {Bloch}}]{Koepsell2020}%
  \BibitemOpen
  \bibfield  {author} {\bibinfo {author} {\bibfnamefont {J.}~\bibnamefont
  {Koepsell}}, \bibinfo {author} {\bibfnamefont {S.}~\bibnamefont {Hirthe}},
  \bibinfo {author} {\bibfnamefont {D.}~\bibnamefont {Bourgund}}, \bibinfo
  {author} {\bibfnamefont {P.}~\bibnamefont {Sompet}}, \bibinfo {author}
  {\bibfnamefont {J.}~\bibnamefont {Vijayan}}, \bibinfo {author} {\bibfnamefont
  {G.}~\bibnamefont {Salomon}}, \bibinfo {author} {\bibfnamefont
  {C.}~\bibnamefont {Gross}}, \ and\ \bibinfo {author} {\bibfnamefont
  {I.}~\bibnamefont {Bloch}},\ }\href {\doibase 10.1103/PhysRevLett.125.010403}
  {\bibfield  {journal} {\bibinfo  {journal} {Phys. Rev. Lett.}\ }\textbf
  {\bibinfo {volume} {125}},\ \bibinfo {pages} {010403} (\bibinfo {year}
  {2020})}\BibitemShut {NoStop}%
\bibitem [{\citenamefont {Hartke}\ \emph {et~al.}(2020)\citenamefont {Hartke},
  \citenamefont {Oreg}, \citenamefont {Jia},\ and\ \citenamefont
  {Zwierlein}}]{Hartke2020}%
  \BibitemOpen
  \bibfield  {author} {\bibinfo {author} {\bibfnamefont {T.}~\bibnamefont
  {Hartke}}, \bibinfo {author} {\bibfnamefont {B.}~\bibnamefont {Oreg}},
  \bibinfo {author} {\bibfnamefont {N.}~\bibnamefont {Jia}}, \ and\ \bibinfo
  {author} {\bibfnamefont {M.}~\bibnamefont {Zwierlein}},\ }\href {\doibase
  10.1103/PhysRevLett.125.113601} {\bibfield  {journal} {\bibinfo  {journal}
  {Phys. Rev. Lett.}\ }\textbf {\bibinfo {volume} {125}},\ \bibinfo {pages}
  {113601} (\bibinfo {year} {2020})}\BibitemShut {NoStop}%
\bibitem [{\citenamefont {Gross}\ and\ \citenamefont {Bakr}(2020)}]{Gross2020}%
  \BibitemOpen
  \bibfield  {author} {\bibinfo {author} {\bibfnamefont {C.}~\bibnamefont
  {Gross}}\ and\ \bibinfo {author} {\bibfnamefont {W.~S.}\ \bibnamefont
  {Bakr}},\ }\href@noop {} {\  (\bibinfo {year} {2020})},\ \Eprint
  {http://arxiv.org/abs/2010.15407} {arXiv:2010.15407} \BibitemShut {NoStop}%
\bibitem [{\citenamefont {Murmann}\ \emph {et~al.}(2015)\citenamefont
  {Murmann}, \citenamefont {Bergschneider}, \citenamefont {Klinkhamer},
  \citenamefont {Z{\"{u}}rn}, \citenamefont {Lompe},\ and\ \citenamefont
  {Jochim}}]{Murmann2015}%
  \BibitemOpen
  \bibfield  {author} {\bibinfo {author} {\bibfnamefont {S.}~\bibnamefont
  {Murmann}}, \bibinfo {author} {\bibfnamefont {A.}~\bibnamefont
  {Bergschneider}}, \bibinfo {author} {\bibfnamefont {V.~M.}\ \bibnamefont
  {Klinkhamer}}, \bibinfo {author} {\bibfnamefont {G.}~\bibnamefont
  {Z{\"{u}}rn}}, \bibinfo {author} {\bibfnamefont {T.}~\bibnamefont {Lompe}}, \
  and\ \bibinfo {author} {\bibfnamefont {S.}~\bibnamefont {Jochim}},\ }\href
  {\doibase 10.1103/PhysRevLett.114.080402} {\bibfield  {journal} {\bibinfo
  {journal} {Phys. Rev. Lett.}\ }\textbf {\bibinfo {volume} {114}},\ \bibinfo
  {pages} {080402} (\bibinfo {year} {2015})}\BibitemShut {NoStop}%
\bibitem [{\citenamefont {Labuhn}\ \emph {et~al.}(2016)\citenamefont {Labuhn},
  \citenamefont {Barredo}, \citenamefont {Ravets}, \citenamefont
  {de~L{\'{e}}s{\'{e}}leuc}, \citenamefont {Macr{\`{i}}}, \citenamefont
  {Lahaye},\ and\ \citenamefont {Browaeys}}]{Labuhn2016Tunable}%
  \BibitemOpen
  \bibfield  {author} {\bibinfo {author} {\bibfnamefont {H.}~\bibnamefont
  {Labuhn}}, \bibinfo {author} {\bibfnamefont {D.}~\bibnamefont {Barredo}},
  \bibinfo {author} {\bibfnamefont {S.}~\bibnamefont {Ravets}}, \bibinfo
  {author} {\bibfnamefont {S.}~\bibnamefont {de~L{\'{e}}s{\'{e}}leuc}},
  \bibinfo {author} {\bibfnamefont {T.}~\bibnamefont {Macr{\`{i}}}}, \bibinfo
  {author} {\bibfnamefont {T.}~\bibnamefont {Lahaye}}, \ and\ \bibinfo {author}
  {\bibfnamefont {A.}~\bibnamefont {Browaeys}},\ }\href {\doibase
  10.1038/nature18274} {\bibfield  {journal} {\bibinfo  {journal} {Nature}\
  }\textbf {\bibinfo {volume} {534}},\ \bibinfo {pages} {667} (\bibinfo {year}
  {2016})}\BibitemShut {NoStop}%
\bibitem [{\citenamefont {Bernien}\ \emph {et~al.}(2017)\citenamefont
  {Bernien}, \citenamefont {Schwartz}, \citenamefont {Keesling}, \citenamefont
  {Levine}, \citenamefont {Omran}, \citenamefont {Pichler}, \citenamefont
  {Choi}, \citenamefont {Zibrov}, \citenamefont {Endres}, \citenamefont
  {Greiner}, \citenamefont {Vuleti{\'{c}}},\ and\ \citenamefont
  {Lukin}}]{Bernien2017Probing}%
  \BibitemOpen
  \bibfield  {author} {\bibinfo {author} {\bibfnamefont {H.}~\bibnamefont
  {Bernien}}, \bibinfo {author} {\bibfnamefont {S.}~\bibnamefont {Schwartz}},
  \bibinfo {author} {\bibfnamefont {A.}~\bibnamefont {Keesling}}, \bibinfo
  {author} {\bibfnamefont {H.}~\bibnamefont {Levine}}, \bibinfo {author}
  {\bibfnamefont {A.}~\bibnamefont {Omran}}, \bibinfo {author} {\bibfnamefont
  {H.}~\bibnamefont {Pichler}}, \bibinfo {author} {\bibfnamefont
  {S.}~\bibnamefont {Choi}}, \bibinfo {author} {\bibfnamefont {A.~S.}\
  \bibnamefont {Zibrov}}, \bibinfo {author} {\bibfnamefont {M.}~\bibnamefont
  {Endres}}, \bibinfo {author} {\bibfnamefont {M.}~\bibnamefont {Greiner}},
  \bibinfo {author} {\bibfnamefont {V.}~\bibnamefont {Vuleti{\'{c}}}}, \ and\
  \bibinfo {author} {\bibfnamefont {M.~D.}\ \bibnamefont {Lukin}},\ }\href
  {\doibase 10.1038/nature24622} {\bibfield  {journal} {\bibinfo  {journal}
  {Nature}\ }\textbf {\bibinfo {volume} {551}},\ \bibinfo {pages} {579}
  (\bibinfo {year} {2017})}\BibitemShut {NoStop}%
\bibitem [{\citenamefont {Mazurenko}\ \emph {et~al.}(2017)\citenamefont
  {Mazurenko}, \citenamefont {Chiu}, \citenamefont {Ji}, \citenamefont
  {Parsons}, \citenamefont {Kan{\'a}sz-Nagy}, \citenamefont {Schmidt},
  \citenamefont {Grusdt}, \citenamefont {Demler}, \citenamefont {Greif},\ and\
  \citenamefont {Greiner}}]{Mazurenko2017Antiferro}%
  \BibitemOpen
  \bibfield  {author} {\bibinfo {author} {\bibfnamefont {A.}~\bibnamefont
  {Mazurenko}}, \bibinfo {author} {\bibfnamefont {C.~S.}\ \bibnamefont {Chiu}},
  \bibinfo {author} {\bibfnamefont {G.}~\bibnamefont {Ji}}, \bibinfo {author}
  {\bibfnamefont {M.~F.}\ \bibnamefont {Parsons}}, \bibinfo {author}
  {\bibfnamefont {M.}~\bibnamefont {Kan{\'a}sz-Nagy}}, \bibinfo {author}
  {\bibfnamefont {R.}~\bibnamefont {Schmidt}}, \bibinfo {author} {\bibfnamefont
  {F.}~\bibnamefont {Grusdt}}, \bibinfo {author} {\bibfnamefont
  {E.}~\bibnamefont {Demler}}, \bibinfo {author} {\bibfnamefont
  {D.}~\bibnamefont {Greif}}, \ and\ \bibinfo {author} {\bibfnamefont
  {M.}~\bibnamefont {Greiner}},\ }\href {\doibase 10.1038/nature22362}
  {\bibfield  {journal} {\bibinfo  {journal} {Nature}\ }\textbf {\bibinfo
  {volume} {545}},\ \bibinfo {pages} {462} (\bibinfo {year}
  {2017})}\BibitemShut {NoStop}%
\bibitem [{\citenamefont {Brown}\ \emph {et~al.}(2019)\citenamefont {Brown},
  \citenamefont {Mitra}, \citenamefont {Guardado-Sanchez}, \citenamefont
  {Nourafkan}, \citenamefont {Reymbaut}, \citenamefont {H{\'e}bert},
  \citenamefont {Bergeron}, \citenamefont {Tremblay}, \citenamefont {Kokalj},
  \citenamefont {Huse}, \citenamefont {Schau{\ss}},\ and\ \citenamefont
  {Bakr}}]{Brown2019Badmetal}%
  \BibitemOpen
  \bibfield  {author} {\bibinfo {author} {\bibfnamefont {P.~T.}\ \bibnamefont
  {Brown}}, \bibinfo {author} {\bibfnamefont {D.}~\bibnamefont {Mitra}},
  \bibinfo {author} {\bibfnamefont {E.}~\bibnamefont {Guardado-Sanchez}},
  \bibinfo {author} {\bibfnamefont {R.}~\bibnamefont {Nourafkan}}, \bibinfo
  {author} {\bibfnamefont {A.}~\bibnamefont {Reymbaut}}, \bibinfo {author}
  {\bibfnamefont {C.-D.}\ \bibnamefont {H{\'e}bert}}, \bibinfo {author}
  {\bibfnamefont {S.}~\bibnamefont {Bergeron}}, \bibinfo {author}
  {\bibfnamefont {A.-M.~S.}\ \bibnamefont {Tremblay}}, \bibinfo {author}
  {\bibfnamefont {J.}~\bibnamefont {Kokalj}}, \bibinfo {author} {\bibfnamefont
  {D.~A.}\ \bibnamefont {Huse}}, \bibinfo {author} {\bibfnamefont
  {P.}~\bibnamefont {Schau{\ss}}}, \ and\ \bibinfo {author} {\bibfnamefont
  {W.~S.}\ \bibnamefont {Bakr}},\ }\href {\doibase 10.1126/science.aat4134}
  {\bibfield  {journal} {\bibinfo  {journal} {Science}\ }\textbf {\bibinfo
  {volume} {363}},\ \bibinfo {pages} {379} (\bibinfo {year}
  {2019})}\BibitemShut {NoStop}%
\bibitem [{\citenamefont {Nichols}\ \emph {et~al.}(2019)\citenamefont
  {Nichols}, \citenamefont {Cheuk}, \citenamefont {Okan}, \citenamefont
  {Hartke}, \citenamefont {Mendez}, \citenamefont {Senthil}, \citenamefont
  {Khatami}, \citenamefont {Zhang},\ and\ \citenamefont
  {Zwierlein}}]{Nichols2019spintransport}%
  \BibitemOpen
  \bibfield  {author} {\bibinfo {author} {\bibfnamefont {M.~A.}\ \bibnamefont
  {Nichols}}, \bibinfo {author} {\bibfnamefont {L.~W.}\ \bibnamefont {Cheuk}},
  \bibinfo {author} {\bibfnamefont {M.}~\bibnamefont {Okan}}, \bibinfo {author}
  {\bibfnamefont {T.~R.}\ \bibnamefont {Hartke}}, \bibinfo {author}
  {\bibfnamefont {E.}~\bibnamefont {Mendez}}, \bibinfo {author} {\bibfnamefont
  {T.}~\bibnamefont {Senthil}}, \bibinfo {author} {\bibfnamefont
  {E.}~\bibnamefont {Khatami}}, \bibinfo {author} {\bibfnamefont
  {H.}~\bibnamefont {Zhang}}, \ and\ \bibinfo {author} {\bibfnamefont {M.~W.}\
  \bibnamefont {Zwierlein}},\ }\href {\doibase 10.1126/science.aat4387}
  {\bibfield  {journal} {\bibinfo  {journal} {Science}\ }\textbf {\bibinfo
  {volume} {363}},\ \bibinfo {pages} {383} (\bibinfo {year}
  {2019})}\BibitemShut {NoStop}%
\bibitem [{\citenamefont {Koepsell}\ \emph {et~al.}(2019)\citenamefont
  {Koepsell}, \citenamefont {Vijayan}, \citenamefont {Sompet}, \citenamefont
  {Grusdt}, \citenamefont {Hilker}, \citenamefont {Demler}, \citenamefont
  {Salomon}, \citenamefont {Bloch},\ and\ \citenamefont
  {Gross}}]{Koepsell2019Polaron}%
  \BibitemOpen
  \bibfield  {author} {\bibinfo {author} {\bibfnamefont {J.}~\bibnamefont
  {Koepsell}}, \bibinfo {author} {\bibfnamefont {J.}~\bibnamefont {Vijayan}},
  \bibinfo {author} {\bibfnamefont {P.}~\bibnamefont {Sompet}}, \bibinfo
  {author} {\bibfnamefont {F.}~\bibnamefont {Grusdt}}, \bibinfo {author}
  {\bibfnamefont {T.~A.}\ \bibnamefont {Hilker}}, \bibinfo {author}
  {\bibfnamefont {E.}~\bibnamefont {Demler}}, \bibinfo {author} {\bibfnamefont
  {G.}~\bibnamefont {Salomon}}, \bibinfo {author} {\bibfnamefont
  {I.}~\bibnamefont {Bloch}}, \ and\ \bibinfo {author} {\bibfnamefont
  {C.}~\bibnamefont {Gross}},\ }\href {\doibase 10.1038/s41586-019-1463-1}
  {\bibfield  {journal} {\bibinfo  {journal} {Nature}\ }\textbf {\bibinfo
  {volume} {572}},\ \bibinfo {pages} {358} (\bibinfo {year}
  {2019})}\BibitemShut {NoStop}%
\bibitem [{\citenamefont {Gall}\ \emph {et~al.}(2021)\citenamefont {Gall},
  \citenamefont {Wurz}, \citenamefont {Samland}, \citenamefont {Chan},\ and\
  \citenamefont {K{\"o}hl}}]{Gall2021competing}%
  \BibitemOpen
  \bibfield  {author} {\bibinfo {author} {\bibfnamefont {M.}~\bibnamefont
  {Gall}}, \bibinfo {author} {\bibfnamefont {N.}~\bibnamefont {Wurz}}, \bibinfo
  {author} {\bibfnamefont {J.}~\bibnamefont {Samland}}, \bibinfo {author}
  {\bibfnamefont {C.~F.}\ \bibnamefont {Chan}}, \ and\ \bibinfo {author}
  {\bibfnamefont {M.}~\bibnamefont {K{\"o}hl}},\ }\href {\doibase
  10.1038/s41586-020-03058-x} {\bibfield  {journal} {\bibinfo  {journal}
  {Nature}\ }\textbf {\bibinfo {volume} {589}},\ \bibinfo {pages} {40}
  (\bibinfo {year} {2021})}\BibitemShut {NoStop}%
\bibitem [{\citenamefont {Viverit}\ \emph {et~al.}(2004)\citenamefont
  {Viverit}, \citenamefont {Menotti}, \citenamefont {Calarco},\ and\
  \citenamefont {Smerzi}}]{Viverit2004}%
  \BibitemOpen
  \bibfield  {author} {\bibinfo {author} {\bibfnamefont {L.}~\bibnamefont
  {Viverit}}, \bibinfo {author} {\bibfnamefont {C.}~\bibnamefont {Menotti}},
  \bibinfo {author} {\bibfnamefont {T.}~\bibnamefont {Calarco}}, \ and\
  \bibinfo {author} {\bibfnamefont {A.}~\bibnamefont {Smerzi}},\ }\href
  {\doibase 10.1103/PhysRevLett.93.110401} {\bibfield  {journal} {\bibinfo
  {journal} {Phys. Rev. Lett.}\ }\textbf {\bibinfo {volume} {93}},\ \bibinfo
  {pages} {110401} (\bibinfo {year} {2004})}\BibitemShut {NoStop}%
\bibitem [{\citenamefont {Serwane}\ \emph {et~al.}(2011)\citenamefont
  {Serwane}, \citenamefont {Zurn}, \citenamefont {Lompe}, \citenamefont
  {Ottenstein}, \citenamefont {Wenz},\ and\ \citenamefont
  {Jochim}}]{Serwane2011}%
  \BibitemOpen
  \bibfield  {author} {\bibinfo {author} {\bibfnamefont {F.}~\bibnamefont
  {Serwane}}, \bibinfo {author} {\bibfnamefont {G.}~\bibnamefont {Zurn}},
  \bibinfo {author} {\bibfnamefont {T.}~\bibnamefont {Lompe}}, \bibinfo
  {author} {\bibfnamefont {T.~B.}\ \bibnamefont {Ottenstein}}, \bibinfo
  {author} {\bibfnamefont {A.~N.}\ \bibnamefont {Wenz}}, \ and\ \bibinfo
  {author} {\bibfnamefont {S.}~\bibnamefont {Jochim}},\ }\href {\doibase
  10.1126/science.1201351} {\bibfield  {journal} {\bibinfo  {journal}
  {Science}\ }\textbf {\bibinfo {volume} {332}},\ \bibinfo {pages} {336}
  (\bibinfo {year} {2011})}\BibitemShut {NoStop}%
\bibitem [{\citenamefont {Chiu}\ \emph {et~al.}(2018)\citenamefont {Chiu},
  \citenamefont {Ji}, \citenamefont {Mazurenko}, \citenamefont {Greif},\ and\
  \citenamefont {Greiner}}]{Chiu2018}%
  \BibitemOpen
  \bibfield  {author} {\bibinfo {author} {\bibfnamefont {C.~S.}\ \bibnamefont
  {Chiu}}, \bibinfo {author} {\bibfnamefont {G.}~\bibnamefont {Ji}}, \bibinfo
  {author} {\bibfnamefont {A.}~\bibnamefont {Mazurenko}}, \bibinfo {author}
  {\bibfnamefont {D.}~\bibnamefont {Greif}}, \ and\ \bibinfo {author}
  {\bibfnamefont {M.}~\bibnamefont {Greiner}},\ }\href {\doibase
  10.1103/PhysRevLett.120.243201} {\bibfield  {journal} {\bibinfo  {journal}
  {Phys. Rev. Lett.}\ }\textbf {\bibinfo {volume} {120}},\ \bibinfo {pages}
  {243201} (\bibinfo {year} {2018})}\BibitemShut {NoStop}%
\bibitem [{\citenamefont {Kwiat}\ \emph {et~al.}(2000)\citenamefont {Kwiat},
  \citenamefont {Berglund}, \citenamefont {Altepeter},\ and\ \citenamefont
  {White}}]{Kwiat2000}%
  \BibitemOpen
  \bibfield  {author} {\bibinfo {author} {\bibfnamefont {P.~G.}\ \bibnamefont
  {Kwiat}}, \bibinfo {author} {\bibfnamefont {A.~J.}\ \bibnamefont {Berglund}},
  \bibinfo {author} {\bibfnamefont {J.~B.}\ \bibnamefont {Altepeter}}, \ and\
  \bibinfo {author} {\bibfnamefont {A.~G.}\ \bibnamefont {White}},\ }\href
  {\doibase 10.1126/science.290.5491.498} {\bibfield  {journal} {\bibinfo
  {journal} {Science}\ }\textbf {\bibinfo {volume} {290}},\ \bibinfo {pages}
  {498} (\bibinfo {year} {2000})}\BibitemShut {NoStop}%
\bibitem [{\citenamefont {Kielpinski}\ \emph {et~al.}(2001)\citenamefont
  {Kielpinski}, \citenamefont {Meyer}, \citenamefont {Rowe}, \citenamefont
  {Sackett}, \citenamefont {Itano}, \citenamefont {Monroe},\ and\ \citenamefont
  {Wineland}}]{Kielpinski2001}%
  \BibitemOpen
  \bibfield  {author} {\bibinfo {author} {\bibfnamefont {D.}~\bibnamefont
  {Kielpinski}}, \bibinfo {author} {\bibfnamefont {V.}~\bibnamefont {Meyer}},
  \bibinfo {author} {\bibfnamefont {M.~A.}\ \bibnamefont {Rowe}}, \bibinfo
  {author} {\bibfnamefont {C.~A.}\ \bibnamefont {Sackett}}, \bibinfo {author}
  {\bibfnamefont {W.~M.}\ \bibnamefont {Itano}}, \bibinfo {author}
  {\bibfnamefont {C.}~\bibnamefont {Monroe}}, \ and\ \bibinfo {author}
  {\bibfnamefont {D.~J.}\ \bibnamefont {Wineland}},\ }\href {\doibase
  10.1126/science.1057357} {\bibfield  {journal} {\bibinfo  {journal}
  {Science}\ }\textbf {\bibinfo {volume} {291}},\ \bibinfo {pages} {1013}
  (\bibinfo {year} {2001})}\BibitemShut {NoStop}%
\bibitem [{\citenamefont {Chin}\ \emph {et~al.}(2010)\citenamefont {Chin},
  \citenamefont {Grimm}, \citenamefont {Julienne},\ and\ \citenamefont
  {Tiesinga}}]{Chin2010Feshbach}%
  \BibitemOpen
  \bibfield  {author} {\bibinfo {author} {\bibfnamefont {C.}~\bibnamefont
  {Chin}}, \bibinfo {author} {\bibfnamefont {R.}~\bibnamefont {Grimm}},
  \bibinfo {author} {\bibfnamefont {P.}~\bibnamefont {Julienne}}, \ and\
  \bibinfo {author} {\bibfnamefont {E.}~\bibnamefont {Tiesinga}},\ }\href
  {\doibase 10.1103/RevModPhys.82.1225} {\bibfield  {journal} {\bibinfo
  {journal} {Rev. Mod. Phys.}\ }\textbf {\bibinfo {volume} {82}},\ \bibinfo
  {pages} {1225} (\bibinfo {year} {2010})}\BibitemShut {NoStop}%
\bibitem [{\citenamefont {Regal}\ \emph {et~al.}(2003)\citenamefont {Regal},
  \citenamefont {Ticknor}, \citenamefont {Bohn},\ and\ \citenamefont
  {Jin}}]{Regal2003Creation}%
  \BibitemOpen
  \bibfield  {author} {\bibinfo {author} {\bibfnamefont {C.~A.}\ \bibnamefont
  {Regal}}, \bibinfo {author} {\bibfnamefont {C.}~\bibnamefont {Ticknor}},
  \bibinfo {author} {\bibfnamefont {J.~L.}\ \bibnamefont {Bohn}}, \ and\
  \bibinfo {author} {\bibfnamefont {D.~S.}\ \bibnamefont {Jin}},\ }\href
  {\doibase 10.1038/nature01738} {\bibfield  {journal} {\bibinfo  {journal}
  {Nature}\ }\textbf {\bibinfo {volume} {424}},\ \bibinfo {pages} {47}
  (\bibinfo {year} {2003})}\BibitemShut {NoStop}%
\bibitem [{\citenamefont {K{\"{o}}hl}\ \emph {et~al.}(2005)\citenamefont
  {K{\"{o}}hl}, \citenamefont {Moritz}, \citenamefont {St{\"{o}}ferle},
  \citenamefont {G{\"{u}}nter},\ and\ \citenamefont {Esslinger}}]{Kohl2005}%
  \BibitemOpen
  \bibfield  {author} {\bibinfo {author} {\bibfnamefont {M.}~\bibnamefont
  {K{\"{o}}hl}}, \bibinfo {author} {\bibfnamefont {H.}~\bibnamefont {Moritz}},
  \bibinfo {author} {\bibfnamefont {T.}~\bibnamefont {St{\"{o}}ferle}},
  \bibinfo {author} {\bibfnamefont {K.}~\bibnamefont {G{\"{u}}nter}}, \ and\
  \bibinfo {author} {\bibfnamefont {T.}~\bibnamefont {Esslinger}},\ }\href
  {\doibase 10.1103/PhysRevLett.94.080403} {\bibfield  {journal} {\bibinfo
  {journal} {Phys. Rev. Lett.}\ }\textbf {\bibinfo {volume} {94}},\ \bibinfo
  {pages} {080403} (\bibinfo {year} {2005})}\BibitemShut {NoStop}%
\bibitem [{\citenamefont {Diener}\ and\ \citenamefont {Ho}(2006)}]{Diener2006}%
  \BibitemOpen
  \bibfield  {author} {\bibinfo {author} {\bibfnamefont {R.~B.}\ \bibnamefont
  {Diener}}\ and\ \bibinfo {author} {\bibfnamefont {T.-L.}\ \bibnamefont
  {Ho}},\ }\href {\doibase 10.1103/PhysRevLett.96.010402} {\bibfield  {journal}
  {\bibinfo  {journal} {Phys. Rev. Lett.}\ }\textbf {\bibinfo {volume} {96}},\
  \bibinfo {pages} {010402} (\bibinfo {year} {2006})}\BibitemShut {NoStop}%
\bibitem [{\citenamefont {Z{\"{u}}rn}\ \emph {et~al.}(2012)\citenamefont
  {Z{\"{u}}rn}, \citenamefont {Serwane}, \citenamefont {Lompe}, \citenamefont
  {Wenz}, \citenamefont {Ries}, \citenamefont {Bohn},\ and\ \citenamefont
  {Jochim}}]{Zurn2012}%
  \BibitemOpen
  \bibfield  {author} {\bibinfo {author} {\bibfnamefont {G.}~\bibnamefont
  {Z{\"{u}}rn}}, \bibinfo {author} {\bibfnamefont {F.}~\bibnamefont {Serwane}},
  \bibinfo {author} {\bibfnamefont {T.}~\bibnamefont {Lompe}}, \bibinfo
  {author} {\bibfnamefont {A.~N.}\ \bibnamefont {Wenz}}, \bibinfo {author}
  {\bibfnamefont {M.~G.}\ \bibnamefont {Ries}}, \bibinfo {author}
  {\bibfnamefont {J.~E.}\ \bibnamefont {Bohn}}, \ and\ \bibinfo {author}
  {\bibfnamefont {S.}~\bibnamefont {Jochim}},\ }\href {\doibase
  10.1103/PhysRevLett.108.075303} {\bibfield  {journal} {\bibinfo  {journal}
  {Phys. Rev. Lett.}\ }\textbf {\bibinfo {volume} {108}},\ \bibinfo {pages}
  {075303} (\bibinfo {year} {2012})}\BibitemShut {NoStop}%
\bibitem [{\citenamefont {Busch}\ \emph {et~al.}(1998)\citenamefont {Busch},
  \citenamefont {Englert}, \citenamefont {Rza{\.{z}}ewski},\ and\ \citenamefont
  {Wilkens}}]{Busch1998}%
  \BibitemOpen
  \bibfield  {author} {\bibinfo {author} {\bibfnamefont {T.}~\bibnamefont
  {Busch}}, \bibinfo {author} {\bibfnamefont {B.-G.}\ \bibnamefont {Englert}},
  \bibinfo {author} {\bibfnamefont {K.}~\bibnamefont {Rza{\.{z}}ewski}}, \ and\
  \bibinfo {author} {\bibfnamefont {M.}~\bibnamefont {Wilkens}},\ }\href
  {\doibase 10.1023/A:1018705520999} {\bibfield  {journal} {\bibinfo  {journal}
  {Found. Phys.}\ }\textbf {\bibinfo {volume} {28}},\ \bibinfo {pages} {549}
  (\bibinfo {year} {1998})}\BibitemShut {NoStop}%
\bibitem [{\citenamefont {Idziaszek}\ and\ \citenamefont
  {Calarco}(2005)}]{Idziaszek2005}%
  \BibitemOpen
  \bibfield  {author} {\bibinfo {author} {\bibfnamefont {Z.}~\bibnamefont
  {Idziaszek}}\ and\ \bibinfo {author} {\bibfnamefont {T.}~\bibnamefont
  {Calarco}},\ }\href {\doibase 10.1103/PhysRevA.71.050701} {\bibfield
  {journal} {\bibinfo  {journal} {Phys. Rev. A}\ }\textbf {\bibinfo {volume}
  {71}},\ \bibinfo {pages} {050701} (\bibinfo {year} {2005})}\BibitemShut
  {NoStop}%
\bibitem [{\citenamefont {Bolda}\ \emph {et~al.}(2005)\citenamefont {Bolda},
  \citenamefont {Tiesinga},\ and\ \citenamefont {Julienne}}]{Bolda2005}%
  \BibitemOpen
  \bibfield  {author} {\bibinfo {author} {\bibfnamefont {E.}~\bibnamefont
  {Bolda}}, \bibinfo {author} {\bibfnamefont {E.}~\bibnamefont {Tiesinga}}, \
  and\ \bibinfo {author} {\bibfnamefont {P.}~\bibnamefont {Julienne}},\ }\href
  {\doibase 10.1103/PhysRevA.71.033404} {\bibfield  {journal} {\bibinfo
  {journal} {Phys. Rev. A}\ }\textbf {\bibinfo {volume} {71}},\ \bibinfo
  {pages} {033404} (\bibinfo {year} {2005})}\BibitemShut {NoStop}%
\bibitem [{\citenamefont {Sala}\ \emph {et~al.}(2013)\citenamefont {Sala},
  \citenamefont {Z{\"{u}}rn}, \citenamefont {Lompe}, \citenamefont {Wenz},
  \citenamefont {Murmann}, \citenamefont {Serwane}, \citenamefont {Jochim},\
  and\ \citenamefont {Saenz}}]{Sala2013}%
  \BibitemOpen
  \bibfield  {author} {\bibinfo {author} {\bibfnamefont {S.}~\bibnamefont
  {Sala}}, \bibinfo {author} {\bibfnamefont {G.}~\bibnamefont {Z{\"{u}}rn}},
  \bibinfo {author} {\bibfnamefont {T.}~\bibnamefont {Lompe}}, \bibinfo
  {author} {\bibfnamefont {A.~N.}\ \bibnamefont {Wenz}}, \bibinfo {author}
  {\bibfnamefont {S.}~\bibnamefont {Murmann}}, \bibinfo {author} {\bibfnamefont
  {F.}~\bibnamefont {Serwane}}, \bibinfo {author} {\bibfnamefont
  {S.}~\bibnamefont {Jochim}}, \ and\ \bibinfo {author} {\bibfnamefont
  {A.}~\bibnamefont {Saenz}},\ }\href {\doibase 10.1103/PhysRevLett.110.203202}
  {\bibfield  {journal} {\bibinfo  {journal} {Phys. Rev. Lett.}\ }\textbf
  {\bibinfo {volume} {110}},\ \bibinfo {pages} {203202} (\bibinfo {year}
  {2013})}\BibitemShut {NoStop}%
\bibitem [{\citenamefont {Sala}\ and\ \citenamefont
  {Saenz}(2016)}]{Sala2016Theory}%
  \BibitemOpen
  \bibfield  {author} {\bibinfo {author} {\bibfnamefont {S.}~\bibnamefont
  {Sala}}\ and\ \bibinfo {author} {\bibfnamefont {A.}~\bibnamefont {Saenz}},\
  }\href {\doibase 10.1103/PhysRevA.94.022713} {\bibfield  {journal} {\bibinfo
  {journal} {Phys. Rev. A}\ }\textbf {\bibinfo {volume} {94}},\ \bibinfo
  {pages} {022713} (\bibinfo {year} {2016})}\BibitemShut {NoStop}%
\bibitem [{\citenamefont {Ishmukhamedov}\ and\ \citenamefont
  {Melezhik}(2017)}]{Ishmukhamedov2017Tunneling}%
  \BibitemOpen
  \bibfield  {author} {\bibinfo {author} {\bibfnamefont {I.~S.}\ \bibnamefont
  {Ishmukhamedov}}\ and\ \bibinfo {author} {\bibfnamefont {V.~S.}\ \bibnamefont
  {Melezhik}},\ }\href {\doibase 10.1103/PhysRevA.95.062701} {\bibfield
  {journal} {\bibinfo  {journal} {Phys. Rev. A}\ }\textbf {\bibinfo {volume}
  {95}},\ \bibinfo {pages} {062701} (\bibinfo {year} {2017})}\BibitemShut
  {NoStop}%
\bibitem [{SI()}]{SI}%
  \BibitemOpen
  \href@noop {} {}\bibinfo {note} {See Supplementary Information.}\BibitemShut
  {Stop}%
\bibitem [{\citenamefont {M{\"{u}}ller}\ \emph {et~al.}(2007)\citenamefont
  {M{\"{u}}ller}, \citenamefont {F{\"{o}}lling}, \citenamefont {Widera},\ and\
  \citenamefont {Bloch}}]{Muller2007}%
  \BibitemOpen
  \bibfield  {author} {\bibinfo {author} {\bibfnamefont {T.}~\bibnamefont
  {M{\"{u}}ller}}, \bibinfo {author} {\bibfnamefont {S.}~\bibnamefont
  {F{\"{o}}lling}}, \bibinfo {author} {\bibfnamefont {A.}~\bibnamefont
  {Widera}}, \ and\ \bibinfo {author} {\bibfnamefont {I.}~\bibnamefont
  {Bloch}},\ }\href {\doibase 10.1103/PhysRevLett.99.200405} {\bibfield
  {journal} {\bibinfo  {journal} {Phys. Rev. Lett.}\ }\textbf {\bibinfo
  {volume} {99}},\ \bibinfo {pages} {200405} (\bibinfo {year}
  {2007})}\BibitemShut {NoStop}%
\bibitem [{\citenamefont {F{\"{o}}rster}\ \emph {et~al.}(2009)\citenamefont
  {F{\"{o}}rster}, \citenamefont {Karski}, \citenamefont {Choi}, \citenamefont
  {Steffen}, \citenamefont {Alt}, \citenamefont {Meschede}, \citenamefont
  {Widera}, \citenamefont {Montano}, \citenamefont {Lee}, \citenamefont
  {Rakreungdet},\ and\ \citenamefont {Jessen}}]{Forster2009}%
  \BibitemOpen
  \bibfield  {author} {\bibinfo {author} {\bibfnamefont {L.}~\bibnamefont
  {F{\"{o}}rster}}, \bibinfo {author} {\bibfnamefont {M.}~\bibnamefont
  {Karski}}, \bibinfo {author} {\bibfnamefont {J.-M.}\ \bibnamefont {Choi}},
  \bibinfo {author} {\bibfnamefont {A.}~\bibnamefont {Steffen}}, \bibinfo
  {author} {\bibfnamefont {W.}~\bibnamefont {Alt}}, \bibinfo {author}
  {\bibfnamefont {D.}~\bibnamefont {Meschede}}, \bibinfo {author}
  {\bibfnamefont {A.}~\bibnamefont {Widera}}, \bibinfo {author} {\bibfnamefont
  {E.}~\bibnamefont {Montano}}, \bibinfo {author} {\bibfnamefont {J.~H.}\
  \bibnamefont {Lee}}, \bibinfo {author} {\bibfnamefont {W.}~\bibnamefont
  {Rakreungdet}}, \ and\ \bibinfo {author} {\bibfnamefont {P.~S.}\ \bibnamefont
  {Jessen}},\ }\href {\doibase 10.1103/PhysRevLett.103.233001} {\bibfield
  {journal} {\bibinfo  {journal} {Phys. Rev. Lett.}\ }\textbf {\bibinfo
  {volume} {103}},\ \bibinfo {pages} {233001} (\bibinfo {year}
  {2009})}\BibitemShut {NoStop}%
\bibitem [{\citenamefont {van Frank}\ \emph {et~al.}(2014)\citenamefont {van
  Frank}, \citenamefont {Negretti}, \citenamefont {Berrada}, \citenamefont
  {B{\"{u}}cker}, \citenamefont {Montangero}, \citenamefont {Schaff},
  \citenamefont {Schumm}, \citenamefont {Calarco},\ and\ \citenamefont
  {Schmiedmayer}}]{VanFrank2014}%
  \BibitemOpen
  \bibfield  {author} {\bibinfo {author} {\bibfnamefont {S.}~\bibnamefont {van
  Frank}}, \bibinfo {author} {\bibfnamefont {A.}~\bibnamefont {Negretti}},
  \bibinfo {author} {\bibfnamefont {T.}~\bibnamefont {Berrada}}, \bibinfo
  {author} {\bibfnamefont {R.}~\bibnamefont {B{\"{u}}cker}}, \bibinfo {author}
  {\bibfnamefont {S.}~\bibnamefont {Montangero}}, \bibinfo {author}
  {\bibfnamefont {J.-F.}\ \bibnamefont {Schaff}}, \bibinfo {author}
  {\bibfnamefont {T.}~\bibnamefont {Schumm}}, \bibinfo {author} {\bibfnamefont
  {T.}~\bibnamefont {Calarco}}, \ and\ \bibinfo {author} {\bibfnamefont
  {J.}~\bibnamefont {Schmiedmayer}},\ }\href {\doibase 10.1038/ncomms5009}
  {\bibfield  {journal} {\bibinfo  {journal} {Nat. Commun.}\ }\textbf {\bibinfo
  {volume} {5}},\ \bibinfo {pages} {4009} (\bibinfo {year} {2014})}\BibitemShut
  {NoStop}%
\bibitem [{\citenamefont {Thompson}\ \emph {et~al.}(2005)\citenamefont
  {Thompson}, \citenamefont {Hodby},\ and\ \citenamefont
  {Wieman}}]{Thompson2005Ultracold}%
  \BibitemOpen
  \bibfield  {author} {\bibinfo {author} {\bibfnamefont {S.~T.}\ \bibnamefont
  {Thompson}}, \bibinfo {author} {\bibfnamefont {E.}~\bibnamefont {Hodby}}, \
  and\ \bibinfo {author} {\bibfnamefont {C.~E.}\ \bibnamefont {Wieman}},\
  }\href {\doibase 10.1103/PhysRevLett.95.190404} {\bibfield  {journal}
  {\bibinfo  {journal} {Phys. Rev. Lett.}\ }\textbf {\bibinfo {volume} {95}},\
  \bibinfo {pages} {190404} (\bibinfo {year} {2005})}\BibitemShut {NoStop}%
\bibitem [{\citenamefont {Fuchs}\ \emph {et~al.}(2009)\citenamefont {Fuchs},
  \citenamefont {Dobrovitski}, \citenamefont {Toyli}, \citenamefont
  {Heremans},\ and\ \citenamefont {Awschalom}}]{Fuchs2009Gigahertz}%
  \BibitemOpen
  \bibfield  {author} {\bibinfo {author} {\bibfnamefont {G.~D.}\ \bibnamefont
  {Fuchs}}, \bibinfo {author} {\bibfnamefont {V.~V.}\ \bibnamefont
  {Dobrovitski}}, \bibinfo {author} {\bibfnamefont {D.~M.}\ \bibnamefont
  {Toyli}}, \bibinfo {author} {\bibfnamefont {F.~J.}\ \bibnamefont {Heremans}},
  \ and\ \bibinfo {author} {\bibfnamefont {D.~D.}\ \bibnamefont {Awschalom}},\
  }\href {\doibase 10.1126/science.1181193} {\bibfield  {journal} {\bibinfo
  {journal} {Science}\ }\textbf {\bibinfo {volume} {326}},\ \bibinfo {pages}
  {1520} (\bibinfo {year} {2009})}\BibitemShut {NoStop}%
\bibitem [{\citenamefont {Schiller}\ \emph {et~al.}(2014)\citenamefont
  {Schiller}, \citenamefont {Bakalov},\ and\ \citenamefont
  {Korobov}}]{Schiller2014Simplest}%
  \BibitemOpen
  \bibfield  {author} {\bibinfo {author} {\bibfnamefont {S.}~\bibnamefont
  {Schiller}}, \bibinfo {author} {\bibfnamefont {D.}~\bibnamefont {Bakalov}}, \
  and\ \bibinfo {author} {\bibfnamefont {V.~I.}\ \bibnamefont {Korobov}},\
  }\href {\doibase 10.1103/PhysRevLett.113.023004} {\bibfield  {journal}
  {\bibinfo  {journal} {Phys. Rev. Lett.}\ }\textbf {\bibinfo {volume} {113}},\
  \bibinfo {pages} {023004} (\bibinfo {year} {2014})}\BibitemShut {NoStop}%
\bibitem [{\citenamefont {Kondov}\ \emph {et~al.}(2019)\citenamefont {Kondov},
  \citenamefont {Lee}, \citenamefont {Leung}, \citenamefont {Liedl},
  \citenamefont {Majewska}, \citenamefont {Moszynski},\ and\ \citenamefont
  {Zelevinsky}}]{Kondov2019Molecular}%
  \BibitemOpen
  \bibfield  {author} {\bibinfo {author} {\bibfnamefont {S.~S.}\ \bibnamefont
  {Kondov}}, \bibinfo {author} {\bibfnamefont {C.-H.}\ \bibnamefont {Lee}},
  \bibinfo {author} {\bibfnamefont {K.~H.}\ \bibnamefont {Leung}}, \bibinfo
  {author} {\bibfnamefont {C.}~\bibnamefont {Liedl}}, \bibinfo {author}
  {\bibfnamefont {I.}~\bibnamefont {Majewska}}, \bibinfo {author}
  {\bibfnamefont {R.}~\bibnamefont {Moszynski}}, \ and\ \bibinfo {author}
  {\bibfnamefont {T.}~\bibnamefont {Zelevinsky}},\ }\href {\doibase
  10.1038/s41567-019-0632-3} {\bibfield  {journal} {\bibinfo  {journal} {Nat.
  Phys.}\ }\textbf {\bibinfo {volume} {15}},\ \bibinfo {pages} {1118} (\bibinfo
  {year} {2019})}\BibitemShut {NoStop}%
\bibitem [{\citenamefont {DeMille}(2002)}]{DeMille2002Quantum}%
  \BibitemOpen
  \bibfield  {author} {\bibinfo {author} {\bibfnamefont {D.}~\bibnamefont
  {DeMille}},\ }\href {\doibase 10.1103/PhysRevLett.88.067901} {\bibfield
  {journal} {\bibinfo  {journal} {Phys. Rev. Lett.}\ }\textbf {\bibinfo
  {volume} {88}},\ \bibinfo {pages} {067901} (\bibinfo {year}
  {2002})}\BibitemShut {NoStop}%
\bibitem [{\citenamefont {Haller}\ \emph {et~al.}(2010)\citenamefont {Haller},
  \citenamefont {Mark}, \citenamefont {Hart}, \citenamefont {Danzl},
  \citenamefont {Reichs{\"{o}}llner}, \citenamefont {Melezhik}, \citenamefont
  {Schmelcher},\ and\ \citenamefont {N{\"{a}}gerl}}]{Haller2010}%
  \BibitemOpen
  \bibfield  {author} {\bibinfo {author} {\bibfnamefont {E.}~\bibnamefont
  {Haller}}, \bibinfo {author} {\bibfnamefont {M.~J.}\ \bibnamefont {Mark}},
  \bibinfo {author} {\bibfnamefont {R.}~\bibnamefont {Hart}}, \bibinfo {author}
  {\bibfnamefont {J.~G.}\ \bibnamefont {Danzl}}, \bibinfo {author}
  {\bibfnamefont {L.}~\bibnamefont {Reichs{\"{o}}llner}}, \bibinfo {author}
  {\bibfnamefont {V.}~\bibnamefont {Melezhik}}, \bibinfo {author}
  {\bibfnamefont {P.}~\bibnamefont {Schmelcher}}, \ and\ \bibinfo {author}
  {\bibfnamefont {H.-C.}\ \bibnamefont {N{\"{a}}gerl}},\ }\href {\doibase
  10.1103/PhysRevLett.104.153203} {\bibfield  {journal} {\bibinfo  {journal}
  {Phys. Rev. Lett.}\ }\textbf {\bibinfo {volume} {104}},\ \bibinfo {pages}
  {153203} (\bibinfo {year} {2010})}\BibitemShut {NoStop}%
\bibitem [{\citenamefont {Donley}\ \emph {et~al.}(2002)\citenamefont {Donley},
  \citenamefont {Claussen}, \citenamefont {Thompson},\ and\ \citenamefont
  {Wieman}}]{Donley2002}%
  \BibitemOpen
  \bibfield  {author} {\bibinfo {author} {\bibfnamefont {E.~A.}\ \bibnamefont
  {Donley}}, \bibinfo {author} {\bibfnamefont {N.~R.}\ \bibnamefont
  {Claussen}}, \bibinfo {author} {\bibfnamefont {S.~T.}\ \bibnamefont
  {Thompson}}, \ and\ \bibinfo {author} {\bibfnamefont {C.~E.}\ \bibnamefont
  {Wieman}},\ }\href {\doibase 10.1038/417529a} {\bibfield  {journal} {\bibinfo
   {journal} {Nature}\ }\textbf {\bibinfo {volume} {417}},\ \bibinfo {pages}
  {529} (\bibinfo {year} {2002})}\BibitemShut {NoStop}%
\bibitem [{\citenamefont {Syassen}\ \emph {et~al.}(2007)\citenamefont
  {Syassen}, \citenamefont {Bauer}, \citenamefont {Lettner}, \citenamefont
  {Dietze}, \citenamefont {Volz}, \citenamefont {D{\"{u}}rr},\ and\
  \citenamefont {Rempe}}]{Syassen2007}%
  \BibitemOpen
  \bibfield  {author} {\bibinfo {author} {\bibfnamefont {N.}~\bibnamefont
  {Syassen}}, \bibinfo {author} {\bibfnamefont {D.~M.}\ \bibnamefont {Bauer}},
  \bibinfo {author} {\bibfnamefont {M.}~\bibnamefont {Lettner}}, \bibinfo
  {author} {\bibfnamefont {D.}~\bibnamefont {Dietze}}, \bibinfo {author}
  {\bibfnamefont {T.}~\bibnamefont {Volz}}, \bibinfo {author} {\bibfnamefont
  {S.}~\bibnamefont {D{\"{u}}rr}}, \ and\ \bibinfo {author} {\bibfnamefont
  {G.}~\bibnamefont {Rempe}},\ }\href {\doibase 10.1103/PhysRevLett.99.033201}
  {\bibfield  {journal} {\bibinfo  {journal} {Phys. Rev. Lett.}\ }\textbf
  {\bibinfo {volume} {99}},\ \bibinfo {pages} {033201} (\bibinfo {year}
  {2007})}\BibitemShut {NoStop}%
\bibitem [{\citenamefont {Weitenberg}\ \emph {et~al.}(2011)\citenamefont
  {Weitenberg}, \citenamefont {Endres}, \citenamefont {Sherson}, \citenamefont
  {Cheneau}, \citenamefont {Schauß}, \citenamefont {Fukuhara}, \citenamefont
  {Bloch},\ and\ \citenamefont {Kuhr}}]{Weitenberg2011Single}%
  \BibitemOpen
  \bibfield  {author} {\bibinfo {author} {\bibfnamefont {C.}~\bibnamefont
  {Weitenberg}}, \bibinfo {author} {\bibfnamefont {M.}~\bibnamefont {Endres}},
  \bibinfo {author} {\bibfnamefont {J.~F.}\ \bibnamefont {Sherson}}, \bibinfo
  {author} {\bibfnamefont {M.}~\bibnamefont {Cheneau}}, \bibinfo {author}
  {\bibfnamefont {P.}~\bibnamefont {Schauß}}, \bibinfo {author} {\bibfnamefont
  {T.}~\bibnamefont {Fukuhara}}, \bibinfo {author} {\bibfnamefont
  {I.}~\bibnamefont {Bloch}}, \ and\ \bibinfo {author} {\bibfnamefont
  {S.}~\bibnamefont {Kuhr}},\ }\href {\doibase 10.1038/nature09827} {\bibfield
  {journal} {\bibinfo  {journal} {Nature}\ }\textbf {\bibinfo {volume} {471}},\
  \bibinfo {pages} {319} (\bibinfo {year} {2011})}\BibitemShut {NoStop}%
\bibitem [{\citenamefont {Hartmann}\ and\ \citenamefont
  {Hahn}(1962)}]{Hartmann1962Nuclear}%
  \BibitemOpen
  \bibfield  {author} {\bibinfo {author} {\bibfnamefont {S.~R.}\ \bibnamefont
  {Hartmann}}\ and\ \bibinfo {author} {\bibfnamefont {E.~L.}\ \bibnamefont
  {Hahn}},\ }\href {\doibase 10.1103/PhysRev.128.2042} {\bibfield  {journal}
  {\bibinfo  {journal} {Phys. Rev.}\ }\textbf {\bibinfo {volume} {128}},\
  \bibinfo {pages} {2042} (\bibinfo {year} {1962})}\BibitemShut {NoStop}%
\bibitem [{\citenamefont {Rabl}\ \emph {et~al.}(2009)\citenamefont {Rabl},
  \citenamefont {Cappellaro}, \citenamefont {Dutt}, \citenamefont {Jiang},
  \citenamefont {Maze},\ and\ \citenamefont {Lukin}}]{Rabl2009Strong}%
  \BibitemOpen
  \bibfield  {author} {\bibinfo {author} {\bibfnamefont {P.}~\bibnamefont
  {Rabl}}, \bibinfo {author} {\bibfnamefont {P.}~\bibnamefont {Cappellaro}},
  \bibinfo {author} {\bibfnamefont {M.~V.~G.}\ \bibnamefont {Dutt}}, \bibinfo
  {author} {\bibfnamefont {L.}~\bibnamefont {Jiang}}, \bibinfo {author}
  {\bibfnamefont {J.~R.}\ \bibnamefont {Maze}}, \ and\ \bibinfo {author}
  {\bibfnamefont {M.~D.}\ \bibnamefont {Lukin}},\ }\href {\doibase
  10.1103/PhysRevB.79.041302} {\bibfield  {journal} {\bibinfo  {journal} {Phys.
  Rev. B}\ }\textbf {\bibinfo {volume} {79}},\ \bibinfo {pages} {041302}
  (\bibinfo {year} {2009})}\BibitemShut {NoStop}%
\bibitem [{\citenamefont {DiVincenzo}(2000)}]{DiVincenzo2000Criteria}%
  \BibitemOpen
  \bibfield  {author} {\bibinfo {author} {\bibfnamefont {D.~P.}\ \bibnamefont
  {DiVincenzo}},\ }\href {\doibase
  10.1002/1521-3978(200009)48:9/11<771::AID-PROP771>3.0.CO;2-E} {\bibfield
  {journal} {\bibinfo  {journal} {Fortschr. Phys.}\ }\textbf {\bibinfo {volume}
  {48}},\ \bibinfo {pages} {771} (\bibinfo {year} {2000})}\BibitemShut
  {NoStop}%
\bibitem [{\citenamefont {Ho}(2006)}]{Ho2006}%
  \BibitemOpen
  \bibfield  {author} {\bibinfo {author} {\bibfnamefont {A.~F.}\ \bibnamefont
  {Ho}},\ }\href {\doibase 10.1103/PhysRevA.73.061601} {\bibfield  {journal}
  {\bibinfo  {journal} {Phys. Rev. A}\ }\textbf {\bibinfo {volume} {73}},\
  \bibinfo {pages} {061601} (\bibinfo {year} {2006})}\BibitemShut {NoStop}%
\bibitem [{\citenamefont {Koga}\ \emph {et~al.}(2004)\citenamefont {Koga},
  \citenamefont {Kawakami}, \citenamefont {Rice},\ and\ \citenamefont
  {Sigrist}}]{Koga2004Orbital}%
  \BibitemOpen
  \bibfield  {author} {\bibinfo {author} {\bibfnamefont {A.}~\bibnamefont
  {Koga}}, \bibinfo {author} {\bibfnamefont {N.}~\bibnamefont {Kawakami}},
  \bibinfo {author} {\bibfnamefont {T.~M.}\ \bibnamefont {Rice}}, \ and\
  \bibinfo {author} {\bibfnamefont {M.}~\bibnamefont {Sigrist}},\ }\href
  {\doibase 10.1103/PhysRevLett.92.216402} {\bibfield  {journal} {\bibinfo
  {journal} {Phys. Rev. Lett.}\ }\textbf {\bibinfo {volume} {92}},\ \bibinfo
  {pages} {216402} (\bibinfo {year} {2004})}\BibitemShut {NoStop}%
\bibitem [{\citenamefont {Kubo}(2007)}]{Kubo2007Pairing}%
  \BibitemOpen
  \bibfield  {author} {\bibinfo {author} {\bibfnamefont {K.}~\bibnamefont
  {Kubo}},\ }\href {\doibase 10.1103/PhysRevB.75.224509} {\bibfield  {journal}
  {\bibinfo  {journal} {Phys. Rev. B}\ }\textbf {\bibinfo {volume} {75}},\
  \bibinfo {pages} {224509} (\bibinfo {year} {2007})}\BibitemShut {NoStop}%
\bibitem [{\citenamefont {Mandel}\ \emph {et~al.}(2003)\citenamefont {Mandel},
  \citenamefont {Greiner}, \citenamefont {Widera}, \citenamefont {Rom},
  \citenamefont {H{\"a}nsch},\ and\ \citenamefont
  {Bloch}}]{Mandel2003controlled}%
  \BibitemOpen
  \bibfield  {author} {\bibinfo {author} {\bibfnamefont {O.}~\bibnamefont
  {Mandel}}, \bibinfo {author} {\bibfnamefont {M.}~\bibnamefont {Greiner}},
  \bibinfo {author} {\bibfnamefont {A.}~\bibnamefont {Widera}}, \bibinfo
  {author} {\bibfnamefont {T.}~\bibnamefont {Rom}}, \bibinfo {author}
  {\bibfnamefont {T.~W.}\ \bibnamefont {H{\"a}nsch}}, \ and\ \bibinfo {author}
  {\bibfnamefont {I.}~\bibnamefont {Bloch}},\ }\href {\doibase
  10.1038/nature02008} {\bibfield  {journal} {\bibinfo  {journal} {Nature}\
  }\textbf {\bibinfo {volume} {425}},\ \bibinfo {pages} {937} (\bibinfo {year}
  {2003})}\BibitemShut {NoStop}%
\bibitem [{\citenamefont {Anderlini}\ \emph {et~al.}(2007)\citenamefont
  {Anderlini}, \citenamefont {Lee}, \citenamefont {Brown}, \citenamefont
  {Sebby-Strabley}, \citenamefont {Phillips},\ and\ \citenamefont
  {Porto}}]{Anderlini2007}%
  \BibitemOpen
  \bibfield  {author} {\bibinfo {author} {\bibfnamefont {M.}~\bibnamefont
  {Anderlini}}, \bibinfo {author} {\bibfnamefont {P.~J.}\ \bibnamefont {Lee}},
  \bibinfo {author} {\bibfnamefont {B.~L.}\ \bibnamefont {Brown}}, \bibinfo
  {author} {\bibfnamefont {J.}~\bibnamefont {Sebby-Strabley}}, \bibinfo
  {author} {\bibfnamefont {W.~D.}\ \bibnamefont {Phillips}}, \ and\ \bibinfo
  {author} {\bibfnamefont {J.~V.}\ \bibnamefont {Porto}},\ }\href {\doibase
  10.1038/nature06011} {\bibfield  {journal} {\bibinfo  {journal} {Nature}\
  }\textbf {\bibinfo {volume} {448}},\ \bibinfo {pages} {452} (\bibinfo {year}
  {2007})}\BibitemShut {NoStop}%
\bibitem [{\citenamefont {Greif}\ \emph {et~al.}(2013)\citenamefont {Greif},
  \citenamefont {Uehlinger}, \citenamefont {Jotzu}, \citenamefont {Tarruell},\
  and\ \citenamefont {Esslinger}}]{Greif2013Short}%
  \BibitemOpen
  \bibfield  {author} {\bibinfo {author} {\bibfnamefont {D.}~\bibnamefont
  {Greif}}, \bibinfo {author} {\bibfnamefont {T.}~\bibnamefont {Uehlinger}},
  \bibinfo {author} {\bibfnamefont {G.}~\bibnamefont {Jotzu}}, \bibinfo
  {author} {\bibfnamefont {L.}~\bibnamefont {Tarruell}}, \ and\ \bibinfo
  {author} {\bibfnamefont {T.}~\bibnamefont {Esslinger}},\ }\href {\doibase
  10.1126/science.1236362} {\bibfield  {journal} {\bibinfo  {journal}
  {Science}\ }\textbf {\bibinfo {volume} {340}},\ \bibinfo {pages} {1307}
  (\bibinfo {year} {2013})},\ \Eprint {http://arxiv.org/abs/1212.2634}
  {1212.2634} \BibitemShut {NoStop}%
\bibitem [{\citenamefont {Mamaev}\ \emph {et~al.}(2020)\citenamefont {Mamaev},
  \citenamefont {Thywissen},\ and\ \citenamefont {Rey}}]{Mamaev2020Quantum}%
  \BibitemOpen
  \bibfield  {author} {\bibinfo {author} {\bibfnamefont {M.}~\bibnamefont
  {Mamaev}}, \bibinfo {author} {\bibfnamefont {J.~H.}\ \bibnamefont
  {Thywissen}}, \ and\ \bibinfo {author} {\bibfnamefont {A.~M.}\ \bibnamefont
  {Rey}},\ }\href {\doibase 10.1002/qute.201900132} {\bibfield  {journal}
  {\bibinfo  {journal} {Adv. Quantum Technol.}\ }\textbf {\bibinfo {volume}
  {3}},\ \bibinfo {pages} {1900132} (\bibinfo {year} {2020})}\BibitemShut
  {NoStop}%
\bibitem [{\citenamefont {Hollerith}\ \emph {et~al.}(2019)\citenamefont
  {Hollerith}, \citenamefont {Zeiher}, \citenamefont {Rui}, \citenamefont
  {Rubio-Abadal}, \citenamefont {Walther}, \citenamefont {Pohl}, \citenamefont
  {Stamper-Kurn}, \citenamefont {Bloch},\ and\ \citenamefont
  {Gross}}]{Hollerith2019}%
  \BibitemOpen
  \bibfield  {author} {\bibinfo {author} {\bibfnamefont {S.}~\bibnamefont
  {Hollerith}}, \bibinfo {author} {\bibfnamefont {J.}~\bibnamefont {Zeiher}},
  \bibinfo {author} {\bibfnamefont {J.}~\bibnamefont {Rui}}, \bibinfo {author}
  {\bibfnamefont {A.}~\bibnamefont {Rubio-Abadal}}, \bibinfo {author}
  {\bibfnamefont {V.}~\bibnamefont {Walther}}, \bibinfo {author} {\bibfnamefont
  {T.}~\bibnamefont {Pohl}}, \bibinfo {author} {\bibfnamefont {D.~M.}\
  \bibnamefont {Stamper-Kurn}}, \bibinfo {author} {\bibfnamefont
  {I.}~\bibnamefont {Bloch}}, \ and\ \bibinfo {author} {\bibfnamefont
  {C.}~\bibnamefont {Gross}},\ }\href {\doibase 10.1126/science.aaw4150}
  {\bibfield  {journal} {\bibinfo  {journal} {Science}\ }\textbf {\bibinfo
  {volume} {364}},\ \bibinfo {pages} {664} (\bibinfo {year}
  {2019})}\BibitemShut {NoStop}%
\bibitem [{\citenamefont {Yan}\ \emph {et~al.}(2013)\citenamefont {Yan},
  \citenamefont {Moses}, \citenamefont {Gadway}, \citenamefont {Covey},
  \citenamefont {Hazzard}, \citenamefont {Rey}, \citenamefont {Jin},\ and\
  \citenamefont {Ye}}]{Yan2013Observation}%
  \BibitemOpen
  \bibfield  {author} {\bibinfo {author} {\bibfnamefont {B.}~\bibnamefont
  {Yan}}, \bibinfo {author} {\bibfnamefont {S.~A.}\ \bibnamefont {Moses}},
  \bibinfo {author} {\bibfnamefont {B.}~\bibnamefont {Gadway}}, \bibinfo
  {author} {\bibfnamefont {J.~P.}\ \bibnamefont {Covey}}, \bibinfo {author}
  {\bibfnamefont {K.~R.~A.}\ \bibnamefont {Hazzard}}, \bibinfo {author}
  {\bibfnamefont {A.~M.}\ \bibnamefont {Rey}}, \bibinfo {author} {\bibfnamefont
  {D.~S.}\ \bibnamefont {Jin}}, \ and\ \bibinfo {author} {\bibfnamefont
  {J.}~\bibnamefont {Ye}},\ }\href {\doibase 10.1038/nature12483} {\bibfield
  {journal} {\bibinfo  {journal} {Nature}\ }\textbf {\bibinfo {volume} {501}},\
  \bibinfo {pages} {521} (\bibinfo {year} {2013})}\BibitemShut {NoStop}%
\bibitem [{\citenamefont {Baron}\ \emph {et~al.}(2014)\citenamefont {Baron},
  \citenamefont {Campbell}, \citenamefont {DeMille}, \citenamefont {Doyle},
  \citenamefont {Gabrielse}, \citenamefont {Gurevich}, \citenamefont {Hess},
  \citenamefont {Hutzler}, \citenamefont {Kirilov}, \citenamefont {Kozyryev},
  \citenamefont {O'Leary}, \citenamefont {Panda}, \citenamefont {Parsons},
  \citenamefont {Petrik}, \citenamefont {Spaun}, \citenamefont {Vutha},\ and\
  \citenamefont {West}}]{Baron2014Order}%
  \BibitemOpen
  \bibfield  {author} {\bibinfo {author} {\bibfnamefont {J.}~\bibnamefont
  {Baron}}, \bibinfo {author} {\bibfnamefont {W.~C.}\ \bibnamefont {Campbell}},
  \bibinfo {author} {\bibfnamefont {D.}~\bibnamefont {DeMille}}, \bibinfo
  {author} {\bibfnamefont {J.~M.}\ \bibnamefont {Doyle}}, \bibinfo {author}
  {\bibfnamefont {G.}~\bibnamefont {Gabrielse}}, \bibinfo {author}
  {\bibfnamefont {Y.~V.}\ \bibnamefont {Gurevich}}, \bibinfo {author}
  {\bibfnamefont {P.~W.}\ \bibnamefont {Hess}}, \bibinfo {author}
  {\bibfnamefont {N.~R.}\ \bibnamefont {Hutzler}}, \bibinfo {author}
  {\bibfnamefont {E.}~\bibnamefont {Kirilov}}, \bibinfo {author} {\bibfnamefont
  {I.}~\bibnamefont {Kozyryev}}, \bibinfo {author} {\bibfnamefont {B.~R.}\
  \bibnamefont {O'Leary}}, \bibinfo {author} {\bibfnamefont {C.~D.}\
  \bibnamefont {Panda}}, \bibinfo {author} {\bibfnamefont {M.~F.}\ \bibnamefont
  {Parsons}}, \bibinfo {author} {\bibfnamefont {E.~S.}\ \bibnamefont {Petrik}},
  \bibinfo {author} {\bibfnamefont {B.}~\bibnamefont {Spaun}}, \bibinfo
  {author} {\bibfnamefont {A.~C.}\ \bibnamefont {Vutha}}, \ and\ \bibinfo
  {author} {\bibfnamefont {A.~D.}\ \bibnamefont {West}},\ }\href {\doibase
  10.1126/science.1248213} {\bibfield  {journal} {\bibinfo  {journal}
  {Science}\ }\textbf {\bibinfo {volume} {343}},\ \bibinfo {pages} {269}
  (\bibinfo {year} {2014})}\BibitemShut {NoStop}%
\bibitem [{\citenamefont {Cairncross}\ \emph {et~al.}(2017)\citenamefont
  {Cairncross}, \citenamefont {Gresh}, \citenamefont {Grau}, \citenamefont
  {Cossel}, \citenamefont {Roussy}, \citenamefont {Ni}, \citenamefont {Zhou},
  \citenamefont {Ye},\ and\ \citenamefont {Cornell}}]{Cairncross2017Precision}%
  \BibitemOpen
  \bibfield  {author} {\bibinfo {author} {\bibfnamefont {W.~B.}\ \bibnamefont
  {Cairncross}}, \bibinfo {author} {\bibfnamefont {D.~N.}\ \bibnamefont
  {Gresh}}, \bibinfo {author} {\bibfnamefont {M.}~\bibnamefont {Grau}},
  \bibinfo {author} {\bibfnamefont {K.~C.}\ \bibnamefont {Cossel}}, \bibinfo
  {author} {\bibfnamefont {T.~S.}\ \bibnamefont {Roussy}}, \bibinfo {author}
  {\bibfnamefont {Y.}~\bibnamefont {Ni}}, \bibinfo {author} {\bibfnamefont
  {Y.}~\bibnamefont {Zhou}}, \bibinfo {author} {\bibfnamefont {J.}~\bibnamefont
  {Ye}}, \ and\ \bibinfo {author} {\bibfnamefont {E.~A.}\ \bibnamefont
  {Cornell}},\ }\href {\doibase 10.1103/PhysRevLett.119.153001} {\bibfield
  {journal} {\bibinfo  {journal} {Phys. Rev. Lett.}\ }\textbf {\bibinfo
  {volume} {119}},\ \bibinfo {pages} {153001} (\bibinfo {year}
  {2017})}\BibitemShut {NoStop}%
\bibitem [{\citenamefont {D'Errico}\ \emph {et~al.}(2007)\citenamefont
  {D'Errico}, \citenamefont {Zaccanti}, \citenamefont {Fattori}, \citenamefont
  {Roati}, \citenamefont {Inguscio}, \citenamefont {Modugno},\ and\
  \citenamefont {Simoni}}]{DErrico2007Feshbach}%
  \BibitemOpen
  \bibfield  {author} {\bibinfo {author} {\bibfnamefont {C.}~\bibnamefont
  {D'Errico}}, \bibinfo {author} {\bibfnamefont {M.}~\bibnamefont {Zaccanti}},
  \bibinfo {author} {\bibfnamefont {M.}~\bibnamefont {Fattori}}, \bibinfo
  {author} {\bibfnamefont {G.}~\bibnamefont {Roati}}, \bibinfo {author}
  {\bibfnamefont {M.}~\bibnamefont {Inguscio}}, \bibinfo {author}
  {\bibfnamefont {G.}~\bibnamefont {Modugno}}, \ and\ \bibinfo {author}
  {\bibfnamefont {A.}~\bibnamefont {Simoni}},\ }\href {\doibase
  10.1088/1367-2630/9/7/223} {\bibfield  {journal} {\bibinfo  {journal} {New J.
  Phys.}\ }\textbf {\bibinfo {volume} {9}},\ \bibinfo {pages} {223} (\bibinfo
  {year} {2007})}\BibitemShut {NoStop}%
\bibitem [{\citenamefont {Mitra}\ \emph {et~al.}(2018)\citenamefont {Mitra},
  \citenamefont {Brown}, \citenamefont {Guardado-Sanchez}, \citenamefont
  {Kondov}, \citenamefont {Devakul}, \citenamefont {Huse}, \citenamefont
  {Schau{\ss}},\ and\ \citenamefont {Bakr}}]{Mitra2018}%
  \BibitemOpen
  \bibfield  {author} {\bibinfo {author} {\bibfnamefont {D.}~\bibnamefont
  {Mitra}}, \bibinfo {author} {\bibfnamefont {P.~T.}\ \bibnamefont {Brown}},
  \bibinfo {author} {\bibfnamefont {E.}~\bibnamefont {Guardado-Sanchez}},
  \bibinfo {author} {\bibfnamefont {S.~S.}\ \bibnamefont {Kondov}}, \bibinfo
  {author} {\bibfnamefont {T.}~\bibnamefont {Devakul}}, \bibinfo {author}
  {\bibfnamefont {D.~A.}\ \bibnamefont {Huse}}, \bibinfo {author}
  {\bibfnamefont {P.}~\bibnamefont {Schau{\ss}}}, \ and\ \bibinfo {author}
  {\bibfnamefont {W.~S.}\ \bibnamefont {Bakr}},\ }\href {\doibase
  10.1038/nphys4297} {\bibfield  {journal} {\bibinfo  {journal} {Nat. Phys.}\
  }\textbf {\bibinfo {volume} {14}},\ \bibinfo {pages} {173} (\bibinfo {year}
  {2018})}\BibitemShut {NoStop}%
\bibitem [{\citenamefont {Chen}\ \emph {et~al.}(2020)\citenamefont {Chen},
  \citenamefont {Xiao}, \citenamefont {Zhang},\ and\ \citenamefont
  {Zhang}}]{Chen2020Analytical}%
  \BibitemOpen
  \bibfield  {author} {\bibinfo {author} {\bibfnamefont {Y.}~\bibnamefont
  {Chen}}, \bibinfo {author} {\bibfnamefont {D.-W.}\ \bibnamefont {Xiao}},
  \bibinfo {author} {\bibfnamefont {R.}~\bibnamefont {Zhang}}, \ and\ \bibinfo
  {author} {\bibfnamefont {P.}~\bibnamefont {Zhang}},\ }\href {\doibase
  10.1103/PhysRevA.101.053624} {\bibfield  {journal} {\bibinfo  {journal}
  {Phys. Rev. A}\ }\textbf {\bibinfo {volume} {101}},\ \bibinfo {pages}
  {053624} (\bibinfo {year} {2020})}\BibitemShut {NoStop}%
\bibitem [{Sim()}]{SimoniFeshbachTheory}%
  \BibitemOpen
  \href@noop {} {}\bibinfo {note} {Private communication, calculations based on
  the model in Ref.~\cite{DErrico2007Feshbach}.}\BibitemShut {Stop}%
\bibitem [{\citenamefont {Daniel}(2020)}]{Daniel2020Exact}%
  \BibitemOpen
  \bibfield  {author} {\bibinfo {author} {\bibfnamefont {D.~J.}\ \bibnamefont
  {Daniel}},\ }\href {\doibase 10.1093/ptep/ptaa024} {\bibfield  {journal}
  {\bibinfo  {journal} {Prog. Theor. Exp. Phys.}\ }\textbf {\bibinfo {volume}
  {2020}},\ \bibinfo {pages} {043A01} (\bibinfo {year} {2020})}\BibitemShut
  {NoStop}%
\bibitem [{\citenamefont {Laucht}\ \emph {et~al.}(2016)\citenamefont {Laucht},
  \citenamefont {Simmons}, \citenamefont {Kalra}, \citenamefont {Tosi},
  \citenamefont {Dehollain}, \citenamefont {Muhonen}, \citenamefont {Freer},
  \citenamefont {Hudson}, \citenamefont {Itoh}, \citenamefont {Jamieson},
  \citenamefont {McCallum}, \citenamefont {Dzurak},\ and\ \citenamefont
  {Morello}}]{Laucht2016Breaking}%
  \BibitemOpen
  \bibfield  {author} {\bibinfo {author} {\bibfnamefont {A.}~\bibnamefont
  {Laucht}}, \bibinfo {author} {\bibfnamefont {S.}~\bibnamefont {Simmons}},
  \bibinfo {author} {\bibfnamefont {R.}~\bibnamefont {Kalra}}, \bibinfo
  {author} {\bibfnamefont {G.}~\bibnamefont {Tosi}}, \bibinfo {author}
  {\bibfnamefont {J.~P.}\ \bibnamefont {Dehollain}}, \bibinfo {author}
  {\bibfnamefont {J.~T.}\ \bibnamefont {Muhonen}}, \bibinfo {author}
  {\bibfnamefont {S.}~\bibnamefont {Freer}}, \bibinfo {author} {\bibfnamefont
  {F.~E.}\ \bibnamefont {Hudson}}, \bibinfo {author} {\bibfnamefont {K.~M.}\
  \bibnamefont {Itoh}}, \bibinfo {author} {\bibfnamefont {D.~N.}\ \bibnamefont
  {Jamieson}}, \bibinfo {author} {\bibfnamefont {J.~C.}\ \bibnamefont
  {McCallum}}, \bibinfo {author} {\bibfnamefont {A.~S.}\ \bibnamefont
  {Dzurak}}, \ and\ \bibinfo {author} {\bibfnamefont {A.}~\bibnamefont
  {Morello}},\ }\href {\doibase 10.1103/PhysRevB.94.161302} {\bibfield
  {journal} {\bibinfo  {journal} {Phys. Rev. B}\ }\textbf {\bibinfo {volume}
  {94}},\ \bibinfo {pages} {161302} (\bibinfo {year} {2016})}\BibitemShut
  {NoStop}%
\bibitem [{\citenamefont {Wang}\ \emph {et~al.}(2021)\citenamefont {Wang},
  \citenamefont {Liu},\ and\ \citenamefont {Cappellaro}}]{Wang2020Observation}%
  \BibitemOpen
  \bibfield  {author} {\bibinfo {author} {\bibfnamefont {G.}~\bibnamefont
  {Wang}}, \bibinfo {author} {\bibfnamefont {Y.-X.}\ \bibnamefont {Liu}}, \
  and\ \bibinfo {author} {\bibfnamefont {P.}~\bibnamefont {Cappellaro}},\
  }\href {\doibase 10.1103/PhysRevA.103.022415} {\bibfield  {journal} {\bibinfo
   {journal} {Phys. Rev. A}\ }\textbf {\bibinfo {volume} {103}},\ \bibinfo
  {pages} {022415} (\bibinfo {year} {2021})}\BibitemShut {NoStop}%
\end{thebibliography}%

\section{Methods}

\subsection{Experimental setup}

Fermion pair qubits are composed of atoms in the two lowest hyperfine states of ${}^{40}$K:  $\ket{F{=}9/2, m_F{=}-9/2}$ and $\ket{F{=}9/2, m_F{=}-7/2}$. These states possess a ${\sim}7\,$G wide $s$-wave Feshbach resonance at $B_{\infty}{\sim}202.1\,$G and a zero crossing of interactions at $B_0{=}209.094(8)\,$G (measured). The spatial potential experienced by each atom is formed by two 1064$\,$nm lattice beams reflecting off a superpolished substrate forming the first facet of the microscope objective~\cite{Cheuk2015Quantum,Hartke2020}. Each beam propagates near-horizontally in either the $x$ or $y$ direction~(see~Fig.~\ref{fig:Fig_Schematic}(a)), with a shallow angle of incidence to the $x$-$y$ plane of ${\sim}10.2^{\circ}$, and is polarized in the $x$-$y$ plane. The two beams reflect off the (horizontal) substrate of the quantum gas microscope at this angle to form a long wavelength lattice in the $z$ direction ($a_z {\approx}3{\,}\mu$m), before being directly retro-reflected to form a short wavelength lattice in the $x$ and $y$ directions ($a_x{\approx}a_y{\approx}541\,$nm). The net potential in the $z$ direction on each lattice site is the sum of the two potentials formed by the two lattice beams. After initialization, the potential is kept sufficiently deep to prevent all tunneling between sites on relevant timescales.

In a typical experiment, atoms are prepared near $151\,$G at repulsive interactions. To initialize fermion pairs in the qubit subspace, the magnetic field is ramped across the Feshbach resonance to $208\,$G in ${\sim}80{\,}$ms, which is sufficiently fast to avoid narrow resonances between fermion pairs and molecules in higher COM states, either in transverse~\cite{Sala2013} or $z$-directional motion, that all occur at fields below ${\sim}202\,$G~\cite{Idziaszek2005}. A further field ramp to zero interactions at $209.094\,$G, in another ${\sim}100\,$ms, initializes fermion pairs in the state $\ket{1,1}$, the upper of the two recoil eigenstates. At the recoil gap, the measured loss rates for the two pair qubit states $\ket{1,1}$ and $\ket{0,2}_{\rm s}$ are $0.00(2)\,$Hz and $0.08(3)\,$Hz, respectively, and the measured bit-flip rates are $0.06(1)\,$Hz and $0.05(3)\,$Hz.

To drive Rabi oscillations between $\ket{1,1}$ and $\ket{0,2}_{\rm s}$ at the recoil gap, interactions are modulated from repulsive to attractive at 140.65$\,$Hz using the magnetic field~(Fig.~\ref{fig:Fig_Rabi}(a))~\cite{SI}. State $\ket{1,1}$ is converted to $\ket{0,2}_{\rm s}$ and vice versa with probability $99.9973(3)\%$, as calculated from the decay of the contrast of Rabi oscillations ($f_{\rm Rabi}{=}23.902(4)\,{\rm Hz}$, $\tau{=}4.0(3)\,$s, Gaussian~fit). Some fermion pairs in excited vibrational states give spurious, but constant, contributions to the signal, reducing oscillation contrast to $92(1)\%$. A route to improve this contrast through better sample preparation techniques has been demonstrated in recent experiments~\cite{Chiu2018} that achieve low-entropy arrays with fermion pair densities exceeding $99.5\%$.

For readout, an additional 1064$\,$nm superlattice which is directly retro-reflected in the $z$ direction (532$\,$nm lattice spacing) coherently separates the long wavelength $z$ lattice into a double well (Fig.~\ref{fig:Fig_Rabi}(b))~\cite{Hartke2020}. This superlattice remains off except during readout. With this method, the full occupancy of each site in the 2D array (0, 1, or 2 atoms) can be distinguished in fluorescence imaging~\cite{Hartke2020}. Here, to more easily distinguish the presence of two atoms on a given site, the relative fluorescence of the two vertical layers is intentionally imbalanced, and the double well is tilted to move all singly-occupied sites into the darker layer. Only lattice sites with two atoms originally in state $\ket{1,1}$ appear bright, while sites with two atoms initially in $\ket{0,2}_{\rm s}$ appear fully dark. The variable $n_{\ket{1,1}}$ counts the bright sites in a circle of radius 10 sites at the center of the atomic cloud. Repeated images of the same cloud reveal an imaging loss of $n_{\ket{1,1}}$ of ${\sim}12\%$. Improved imaging loss rates near $\sim$2\% have been demonstrated~\cite{Mitra2018}.

\subsection{Energy spectrum calculation}
The energy spectrum of two identical atoms in a 3D, anisotropic harmonic trap interacting via a delta function potential can be calculated exactly, given the 3D scattering length and the harmonic trap frequencies~\cite{Idziaszek2005, Chen2020Analytical}. The trap frequencies $\omega_x/2\pi{=}96.84(4)\,$kHz, $\omega_y/2\pi{=}96.55(4)\,$kHz, and $\omega_z/2\pi{=}25.09(4)\,$kHz are measured via lattice intensity modulation spectroscopy. The 3D scattering length $a_{\rm 3D}$ of the lowest two hyperfine states of ${}^{40}$K as a function of magnetic field is provided by a theoretical calculation~\cite{SimoniFeshbachTheory}, adjusted for the new precise measurement reported here~(Fig.~\ref{fig:Fig_Ramsey}(c)~inset) on the location of the scattering length zero: $B_0{=}209.094(8)\,$G. The scattering length is well approximated by the formula $a_{\rm 3D}{=}a_{\rm bg}(1 {-} (B_0{-}B_\infty)/(B{-}B_\infty))$ with $B_\infty{=}202.1\,$G and $a_{\rm bg}{=}167.6\,a_0$, where $a_0$ is the Bohr radius. The mean value of $\omega_x$ and $\omega_y$, which differ by less than $0.4\%$, is used to obtain a trap ratio $\omega_{x}/\omega_z{=}\omega_{y}/\omega_z{=}3.853(6)$ for input to the theory for an anisotropic 3D harmonic trap in Fig.~\ref{fig:Fig_Ramsey}(a)~\cite{Idziaszek2005}. This energy spectrum is also used to calculate the energy difference $|\Delta U/2|$ for display in Fig.~\ref{fig:Fig_Rabi}(d). For the theoretical energy difference in Fig.~\ref{fig:Fig_Ramsey}(c) (main panel), the measured recoil gap frequency of 140.76(3)$\,$Hz at zero interactions is added as a Rabi coupling to the spectrum of Fig.~\ref{fig:Fig_Ramsey}(a), likewise for other lattice depths reported in the inset.

\clearpage

%========================================
% Supplemental materials
%========================================
%%TC:ignore
\setcounter{equation}{0}
\setcounter{figure}{0}
\setcounter{secnumdepth}{2}
\renewcommand{\theequation}{S\arabic{equation}}
\renewcommand{\thefigure}{S\arabic{figure}}
\renewcommand{\tocname}{Supplementary Materials}
\renewcommand{\appendixname}{Supplement}

% \tableofcontents
% \appendix

\section*{Supplementary Information}

\subsection{Recoil gap calculation}

The magnitude of the recoil gap at vanishing interactions in a 1D lattice potential can be derived using perturbation theory. Corrections proportional to $z^4$ in the potential are included up to second order in perturbation theory, and corrections proportional to $z^6$ are included to first order in perturbation theory. The resulting energy of vibrational state $|n\rangle$ in a lattice potential $V E_R \sin^2(\pi z/a_z)$ to this order is 
\begin{linenomath} % Makes lineno not number this line
\begin{multline}
E_n=  2E_R \sqrt{V} \left(n + \frac{1}{2}\right) -E_R\left( \frac{2n^2 + 2n +1}{4}\right)
\\
-E_R\frac{1}{\sqrt{V}}\left(\frac{2n^3 + 3n^2 +3n +1}{16}\right). 
\end{multline}
\end{linenomath}
The recoil gap between the pair states $\ket{1,1}$ and $\ket{0,2}_{\rm s}$ is 
\begin{equation}
2E_1 - (E_2 + E_0) = E_R\left(1 + \frac{9}{8 \sqrt{V}}\right).
\end{equation}
Exact energies can be obtained from solutions to Mathieu's equation~\cite{Daniel2020Exact}. 

\subsection{Interacting energy gap calculation}

The two atoms interact via a delta function potential $\hat{U}=(4 \pi\hbar^2 a_{\rm 3D}/m) \delta^{(3)}(\textbf{r}_1 {-}\textbf{r}_2)$~\cite{Idziaszek2005,Busch1998}. The first order perturbative energy shift $\langle \hat{U} \rangle$ for a given state is more easily evaluated in the basis of suitably normalized center-of-mass and relative coordinates, $\textbf{R} \equiv (\textbf{r}_1 {+}\textbf{r}_2)/\sqrt{2}$ and $\textbf{r} \equiv (\textbf{r}_1 {-}\textbf{r}_2)/\sqrt{2}$, respectively. 

We first calculate $\langle \hat{U} \rangle$ for the ground state.
At vanishing interactions, in the absence of anharmonicity, the ground state of the trap is the harmonic oscillator ground state in $x$, $y$, and $z$ for both atoms,
\begin{equation}
    \Psi(\textbf{r}_1,\textbf{r}_2)=
    \prod_{\mu\in \{1,2\},\, i \in \{x,y,z\}}
    \frac{ e^{-(r_{\mu,i}/l_i)^2/2}}{(\pi l_i^2)^{ \frac{1}{4} } }, 
\end{equation}
where $l_i = \sqrt{\hbar/m\omega_i}$ is the harmonic oscillator length for coordinate $r_{\mu,i}$. 
To evaluate $\langle \hat{U} \rangle$, one can transform to $\textbf{R}$ and $\textbf{r}$ coordinates. The form of the wavefunction is identical, 
\begin{equation}
    \Psi(\textbf{R},\textbf{r})=
    \prod_{i \in \{x,y,z\}}
    \frac{ e^{-(R_{i}/l_i)^2/2}}{(\pi l_i^2)^{ \frac{1}{4} } }
    \frac{ e^{-(r_{i}/l_i)^2/2}}{(\pi l_i^2)^{ \frac{1}{4} } },
\end{equation}
and the form of $\hat{U}$ is $(4 \pi\hbar^2 a_{\rm 3D}/m)( \delta^{(3)}(\textbf{r})/2^{3/2})$. The operator $\hat{U}$ does not affect the coordinate $\textbf{R}$, and the wavefunctions of  $\textbf{R}$ are already normalized. Thus the ground state energy shift at weak interactions is 
\begin{equation}
    U\equiv \langle \hat{U} \rangle = 
    \frac{ 4 \pi\hbar^2 a_{\rm 3D}}{m} \frac{1}{2^{3/2}} \frac{1}{\pi^{3/2} l_x l_y l_z }.
\end{equation}

In a quasi-1D geometry, the transverse wavefunction (i.e.~for motion along $x$ and $y$) remains in the harmonic ground state for weak interactions.
One can then work in the basis of normalized harmonic oscillator states of the rotated $z$ coordinates, $z_c{=}(z_1{+}z_2)/\sqrt{2}$ and $z_r{=}(z_1{-}z_2)/\sqrt{2}$, with the understanding that the transverse state is the ground state. The two pair qubit states can be written as $|2\rangle_{\rm COM}|0\rangle_{\rm rel}{=} (\ket{0,2}_{\rm s}{+}\ket{1,1})/\sqrt{2}$ and $|0\rangle_{\rm COM}|2\rangle_{\rm rel}{=}(\ket{0,2}_{\rm s}{-}\ket{1,1})/\sqrt{2}$.
The interaction energy shift depends solely on the magnitude of the relative wavefunction at $z_{\rm rel}{=}0$, which is a factor of $\sqrt{2}$ smaller in state $|2\rangle_{\rm rel}$ than in state $|0\rangle_{\rm rel}$. Thus the energy shift is $U/2$ for $|0\rangle_{\rm COM}|2\rangle_{\rm rel}$, while it is $U$ for $|2\rangle_{\rm COM}|0\rangle_{\rm rel}$.

\begin{figure*}[!t]
	\centering
	\includegraphics[width=\textwidth]{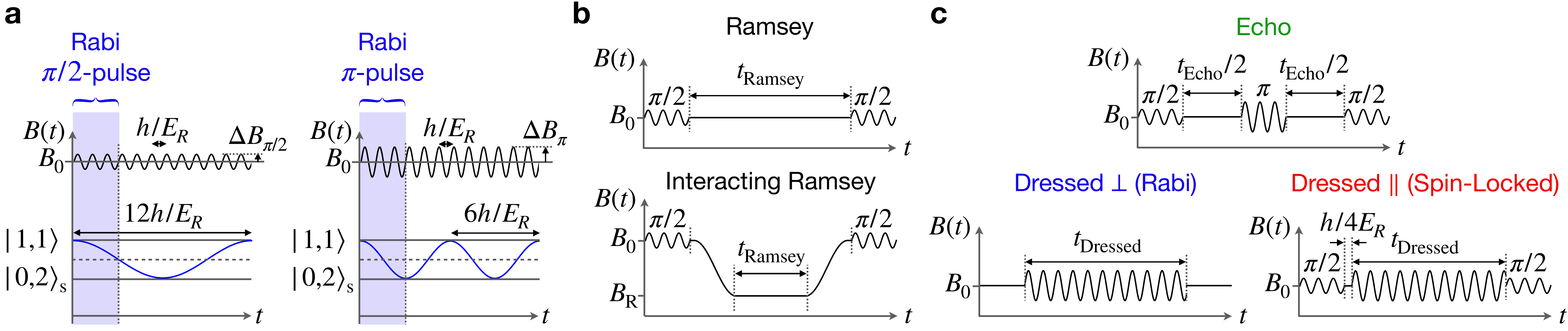}
	% \internallinenumbers
	\caption{\textbf{Qubit control protocols.} 
	(a)~Protocol to transfer population between the fermion pair qubit eigenstates at the recoil gap via a Rabi drive of interactions using the magnetic field~(Fig.~\ref{fig:Fig_Rabi} data). 
	(b)~Protocols for Ramsey measurements of the qubit energy splitting $|\Delta E|$~(Fig.~\ref{fig:Fig_Ramsey} data). 
	(c)~Protocols for measuring coherence at the recoil gap~(Fig.~\ref{fig:Fig_Coherence} data).
	}
	\label{fig:Fig_DriveSchemes}
\end{figure*}

\begin{figure}[!t]
	\centering
	\includegraphics[width=\columnwidth]{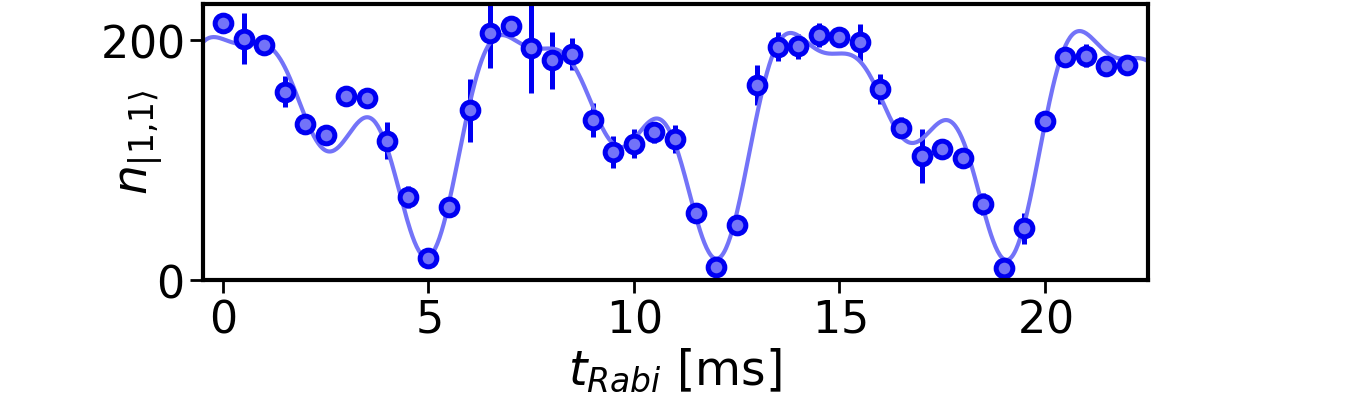}
	% \internallinenumbers
	\caption{\textbf{Strong driving.} 
	A strongly driven Rabi oscillation at the avoided crossing of Fig.~\ref{fig:Fig_Schematic}(c) exhibits non-sinusoidal response. The predicted Rabi coupling $\Delta U/4 {=}h{\times}151.98\,$Hz (see~Fig.~\ref{fig:Fig_Rabi}(d)), which is driven at a modulation frequency of $140.65\,$Hz, is comparable to the recoil energy gap $E_R{=}h{\times}140.76(3)\,$Hz. The solid line shows a phenomenological guide to the eye composed of three sinusoids with frequencies near $E_R/h$, $2E_R/h$, and $3E_R/h$.
	}
	\label{fig:Fig_StrongDriving}
\end{figure}

\begin{figure}[!t]
	\centering
	\includegraphics[width=\columnwidth]{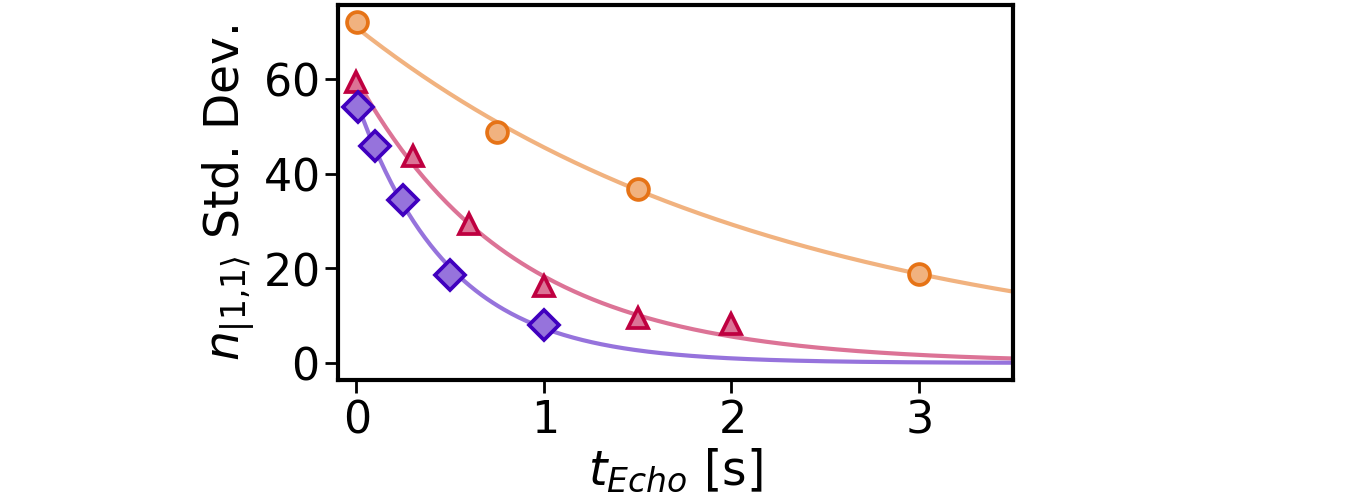}
	% \internallinenumbers
	\caption{\textbf{Coherence at strong interactions.} 
	The standard deviation of the fermion pair qubit state in an echo sequence with randomized extra phase (standard deviation of $n_{\ket{1,1}}$) quantifies the coherence of the register array at strong interactions. A fitted exponential without offset has $1/e$ time constant $\tau{=}2.3(1)\,$s at $|\Delta E|{=}h{\times}1.594(7)\,$kHz (orange), $\tau{=}0.84(5)\,$s at $|\Delta E|{=}h{\times}8.98(7)\,$kHz (red), and $\tau{=}0.49(2)\,$s at $|\Delta E|{=}h{\times}50.7(4)\,$kHz (purple), corresponding to magnetic fields $B{=}206.976(8)\,$G, $B{=}204.235(8)\,$G, and $B{=}202.091(8)\,$G, respectively. Error bars of $\tau$ represent fit error. This data is used in Fig.~\ref{fig:Fig_Coherence}(b).
	}
	\label{fig:Fig_Echo}
\end{figure}

\subsection{Qubit control protocols}
Fig.~\ref{fig:Fig_DriveSchemes} describes in more detail the experimental methods used to control the fermion pair qubit by tuning the interaction strength via the magnetic field. 

To achieve population transfer between $\ket{1,1}$ and $\ket{0,2}_{\rm s}$~(Fig.~\ref{fig:Fig_Rabi} data), pairs are initialized at the recoil gap. Aside from an overall offset, the static effective Hamiltonian near the recoil gap is
\begin{equation}
H = \frac{E_R}{2}\sigma_z+ \frac{U}{4}\sigma_x
\end{equation}
in the basis $(\ket{1,1},\ket{0,2}_{\rm s})$, where $\sigma_z$ and $\sigma_x$ are the Pauli matrices. A Rabi oscillation is driven~(Fig.~\ref{fig:Fig_DriveSchemes}(a)) by modulating the interaction energy $U$ by $\Delta U$ about vanishing interactions at the recoil gap frequency, 
\begin{equation}
H(t) = \frac{E_R}{2}\sigma_z
+\frac{\Delta U}{4}\sin(E_R t/\hbar) \sigma_x.
\end{equation}
This Hamiltonian is produced by sinusoidal modulation of the magnetic field, $B(t) {=} B_0 {+}\Delta B \sin\left( E_R t/\hbar\right)$, with $B_0{=}209.094\,$G being the field where the interactions vanish. The resulting frequency of Rabi oscillation is $f_{\rm Rabi}{=}\Delta U/4h$. Each drive operation begins as a sine wave, and the applied amplitude $\Delta B$ is varied to achieve a $\pi/2$-pulse (left panel of Fig.~\ref{fig:Fig_DriveSchemes}(a), $f_{\rm Rabi}{=}E_R/12 h$, $\Delta B_{\rm \pi/2} {\approx} 50\,$mG) or $\pi$-pulse (right panel of Fig.~\ref{fig:Fig_DriveSchemes}(a), $f_{\rm Rabi}{=} E_R/6 h$, $\Delta B_{\rm \pi} {\approx} 100\,$mG) in exactly 3 drive cycles. 

In a Ramsey measurement~(Fig.~\ref{fig:Fig_Ramsey} data), a $\pi/2$-pulse creates a superposition of $\ket{1,1}$ and $\ket{0,2}_{\rm s}$, the two states acquire a relative phase due to their energy difference, and a $\pi/2$-pulse re-interferes the superposition~(Fig.~\ref{fig:Fig_DriveSchemes}(b), upper panel). A Ramsey measurement at finite interactions proceeds identically, with additional adiabatic magnetic field ramps to and from a field $B_{\rm R}$, where the phase evolution occurs~(lower panel). 

Protocols for measuring coherence~(Fig.~\ref{fig:Fig_Coherence} data) are shown in Fig.~\ref{fig:Fig_DriveSchemes}(c). A $\pi$-pulse is added to the midpoint of the Ramsey protocol to perform an echo sequence~(upper panel). Echo coherence measurements at finite interactions proceed identically, with two additional sets of field ramps to allow phase evolution at a desired field. In the $\perp$ dressed state protocol~(lower left panel), a Rabi drive is continually applied at the $\pi$-pulse amplitude.  In the $\parallel$ dressed state protocol~(lower right panel), a $\pi/2$-pulse is followed by a delay by a quarter of the recoil gap period, effectively shifting the phase of the subsequently applied dressing drive to be aligned with the qubit.

\subsection{Strong driving}
Fig.~\ref{fig:Fig_StrongDriving} shows the fermion pair qubit dynamics under strong driving, with a Rabi coupling comparable to the energy splitting of the effective two level system. The observed multiple frequencies in the response are analogous to the transition frequencies between eigenstates of a two-level atom dressed by light at strong driving. Initializing a specific state where the atom and field are not coupled prepares a superposition of more than two eigenstates of the strongly-coupled system, leading to multi-frequency interference~\cite{Laucht2016Breaking,Wang2020Observation}.

\subsection{Coherence at strong interactions}
An echo sequence (see schematic in Fig.~\ref{fig:Fig_Coherence}(a)) measures the decay of fermion pair qubit coherence at strong interactions (Fig.~\ref{fig:Fig_Echo}). For each repeated experiment at a fixed total time $t_{\rm Echo}$, a random extra phase from 0 to $2\pi$ is added to the fermion pair qubit by redistributing evolution time before and after the echo $\pi$-pulse. The standard deviation of the signal $n_{\ket{1,1}}$ then decays due to loss of coherence within the register array, from which the coherence time is obtained, shown in Fig.~\ref{fig:Fig_Coherence}(b).

\end{document}